\documentclass[acmtocl]{acmtrans2m_calc}

\usepackage{amssymb}
\usepackage{gastex}

\newtheorem{theorem}{Theorem}[section]
\newtheorem{corollary}[theorem]{Corollary}

\newdef{definition}[theorem]{Definition}
\newdef{remark}[theorem]{Remark}
\newdef{example}[theorem]{Example}

\newcommand{\tuple}[1]{\langle #1 \rangle}

\newcommand{\amap}{f}

\newcommand{\aset}{X}
\newcommand{\asetbis}{Y}
\newcommand{\asetter}{Z}

\newcommand{\length}[1]{|#1|}

\newcommand{\aalphabet}{\Sigma}
\newcommand{\aletter}{a}
\newcommand{\aletterbis}{b}

\newcommand{\adataword}{\sigma}
\newcommand{\aword}{w}
\newcommand{\emptyword}{\varepsilon}

\newcommand{\aformula}{\phi}
\newcommand{\aformulabis}{\psi}
\newcommand{\aformulater}{\chi}

\newcommand{\atempop}{\mathtt{O}}
\newcommand{\nextt}{\mathtt{X}}
\newcommand{\until}{\mathtt{U}}
\newcommand{\since}{\mathtt{U}^{-1}}
\newcommand{\sometimes}{\mathtt{F}}
\newcommand{\pastsometimes}{\mathtt{F}^{-1}}
\newcommand{\always}{\mathtt{G}}
\newcommand{\pastalways}{\mathtt{G}^{-1}}

\newcommand{\nuparrow}{\not\,\uparrow}

\newcommand{\eexists}[2]{\exists #1 (#2)}
\newcommand{\fforall}[2]{\forall #1 (#2)}

\newcommand{\agame}{G}
\newcommand{\poss}{P}
\newcommand{\apos}{p}
\newcommand{\aplay}{\pi}
\newcommand{\astr}{\tau}
\newcommand{\ass}{\alpha}
\newcommand{\winn}{W}

\newcommand{\aregaut}{\mathcal{A}}
\newcommand{\aregautbis}{\mathcal{B}}
\newcommand{\aregval}{v}
\newcommand{\acaut}{\mathcal{C}}
\newcommand{\acval}{v}
\newcommand{\amachine}{\mathcal{M}}

\newcommand{\locs}{Q}
\newcommand{\aloc}{q}
\newcommand{\atransf}{\varphi}
\newcommand{\rank}{\rho}
\newcommand{\height}{\gamma}
\newcommand{\ifte}[3]{#2 <\!\!\!\mid\, #1 >\!\!\!\mid\, #3}
\newcommand{\Succ}[1]{\langle\!\langle #1 \rangle\!\rangle}

\newcommand{\egdef}{\stackrel{\mbox{\tiny def}}{=}}

\newcommand{\equivdef}{\stackrel{\mbox{\tiny def}}{\Leftrightarrow}}

\acmVolume{V}
\acmNumber{N}
\acmYear{YY}
\acmMonth{M}

\title
{LTL with the Freeze Quantifier \\
 and Register Automata}

\author
{ST\'EPHANE DEMRI \\
 LSV, CNRS \& ENS Cachan \& INRIA Futurs, France
 \and
 RANKO LAZI\'C \\
 Department of Computer Science, University of Warwick, UK}

\begin{abstract}
A data word is a sequence of pairs of
a letter from a finite alphabet and
an element from an infinite set,
where the latter can only be compared for equality.
To reason about data words,
linear temporal logic is extended by the freeze quantifier,
which stores the element at the current word position into a register,
for equality comparisons deeper in the formula.
By translations from the logic to alternating automata with registers
and then to faulty counter automata
whose counters may erroneously increase at any time,
and from faulty and error-free counter automata to the logic,
we obtain a complete complexity table for logical fragments defined by
varying the set of temporal operators and the number of registers.
In particular, the logic with future-time operators and $1$ register
is decidable but not primitive recursive over finite data words.
Adding past-time operators or $1$ more register,
or switching to infinite data words, cause undecidability.
\end{abstract}

\category
{F.4.1}
{Mathematical Logic and Formal Languages}
{Mathematical Logic}
[Temporal logic]

\category
{F.4.1}
{Mathematical Logic and Formal Languages}
{Formal Languages}
[Decision problems]

\category
{F.1.1}
{Computation by Abstract Devices}
{Models of Computation}
[Automata]

\terms{Algorithms, Verification}

\keywords{Computational complexity, Expressiveness}

\begin{document}

\begin{bottomstuff}
This paper is a revised and extended version of \cite{Demri&Lazic06}.
%\newline
%Authors' e-mail addresses:
%\texttt{demri@lsv.ens-cachan.fr},
%\texttt{lazic@dcs.warwick.ac.uk}.
\newline
The first author was supported
by the ACI ``S\'ecurit\'e et Informatique'' \textsc{Cortos}.
\newline
The second author was supported
by grants from the EPSRC (GR/S52759/01) and the Intel Corporation,
and by ENS Cachan.
\end{bottomstuff}

\maketitle

\section{Introduction}

\subsubsection*{Context}

Logics and automata for words and trees over finite alphabets
are relatively well-understood.  Motivated partly by
the need for formal verification and synthesis of infinite-state systems,
and the search for automated reasoning techniques for XML,
there is an active and broad research programme on
logics and automata for words and trees which have richer structure.

Segoufin's recent survey \cite{Segoufin06} summarises
the substantial progress made on reasoning about data words and data trees.
A data word is a word over a finite alphabet,
with an equivalence relation on word positions.
Implicitly, every word position is labelled by an element (``datum'')
from an infinite set (``data domain''),
but since the infinite set is equipped only with the equality predicate,
it suffices to know which word positions are labelled by equal data,
and that is what the equivalence relation represents.
Similarly, a data tree is a tree (countable, unranked and ordered)
whose every node is labelled by a letter from a finite alphabet,
with an equivalence relation on the set of its nodes.

First-order logic for data words was considered in \cite{Bojanczyketal06a},
where variables range over word positions
($\{0, \ldots, l - 1\}$ or $\mathbb{N}$),
there is a unary predicate for each letter from the finite alphabet,
and there is a binary predicate $x \sim y$ for
the equivalence relation representing equality of data labels.
FO$^2(\sim, <, +1)$ denotes such a logic with two variables
and binary predicates $x + 1 = y$ and $x < y$.
Over finite and over infinite data words,
satisfiability for FO$^2(\sim, <, +1)$ was proved decidable
and at least as hard as reachability for Petri nets.
The latter problem is \textsc{ExpSpace}-hard \cite{Lipton76},
but whether it is elementary has been open for many years.
If the logic is extended by one more variable,
$+1$ becomes expressible using $<$,
but satisfiability was shown undecidable.

Words which contain data from domains with more than the equality predicate
were proposed in \cite{Bouajjanietal07} as models of
configurations of systems with unbounded control structures.
Decidability of satisfiability was proved for
the $\exists^* \forall^*$ fragment of a first-order logic over such words
provided that the underlying logic on data is decidable.

Alternatively to first-order logic over data words,
expressiveness and algorithmic properties of
formalisms based on linear temporal logic were studied in
\cite{French03,Lisitsa&Potapov05,Demri&Lazic&Nowak07,%
Lazic06,Demri&DSouza&Gascon07,Demri&Lazic&Sangnier08}.
LTL was extended by the freeze quantifier:
$\downarrow_r$ stores in register $r$
the equivalence class of the current word position,
and the atomic formula $\uparrow_r$ in its scope is true at a word position
iff the latter belongs to the equivalence class stored in $r$.
Thus, data at different word positions can be compared for equality.
Freeze quantification has also been considered in
timed logics (cf.\ e.g.\ \cite{Alur&Henzinger94}) and
hybrid logics (cf.\ e.g.\ \cite{Goranko96}),
and Fitting has called for an investigation of effects of its addition to
modal logics \cite{Fitting02}.
Let LTL$^\downarrow_n(\mathcal{O})$ denote
LTL with the freeze quantifier, $n$ registers,
and temporal operators $\mathcal{O}$.
Satisfiablity over infinite data words was shown
highly undecidable (i.e., $\Sigma^1_1$-hard)
for LTL$^\downarrow_2(\nextt, \nextt^{-1}, \sometimes, \pastsometimes)$
in \cite{French03} (where $\nextt^{-1}$ and $\pastsometimes$
are the past-time versions of $\nextt$ and $\sometimes$) and
for LTL$^\downarrow_2(\nextt, \until)$
in \cite{Lisitsa&Potapov05,Demri&Lazic&Nowak07},
so complexity (even decidability)
of fragments with $1$ register
remained unknown.
To obtain decidability, various structural restrictions were employed:
flat formulae \cite{Demri&Lazic&Nowak07},
Boolean combinations of safety formulae \cite{Lazic06}, and
that the freeze quantifier is used only for expressing that
the current datum occurs eventually in the future or past
\cite{Demri&DSouza&Gascon07}.
In \cite{Demri&Lazic&Sangnier08}, decidability was obtained
by replacing satisfiability with model checking data words
generated by deterministic one-counter automata.

A third approach to reasoning about data words are register automata
\cite{Kaminski&Francez94,Sakamoto&Ikeda00,Neven&Schwentick&Vianu04}.
In addition to a finite number of control locations,
such an automaton has a finite number of registers
which can store data for later equality comparisons.
In pursuit of a satisfactory notion of regular languages of finite data words,
nonemptiness was shown decidable for one-way nondeterministic register automata,
but the class turned out not to be closed under complement,
and the nonuniversality problem to be undecidable.
However, for such automata $\aregaut$ and $\aregaut'$,
whether the language of $\aregaut$ contains the language of $\aregaut'$
was proved decidable provided $\aregaut$ has only $1$ register
(in the terminology of this paper).
The subclass with $1$ register was thus the best candidate found
for defining regularity, but it is also not closed under complement.
The case of infinite data words was not considered.

\subsubsection*{Contribution}

The main technical achievement of the paper are
translations as depicted in Figure~\ref{f:circle}.
Over finite and over infinite data words,
the translation from the logic to register automata preserves languages,
and for the translations to and from counter automata,
appropriate projections of data words are taken.
For infinite data words, the register automata
have the weak acceptance mechanism \cite{Muller&Saoudi&Schupp86},
which makes them closed under complement.
Alternation is needed because there exist properties
which are expressible in LTL$^\downarrow_1(\nextt, \sometimes)$
(e.g., `no two word positions have equal data')
but not by any one-way nondeterministic register automaton.
The counter automata are one-way, nondeterministic,
accept infinite words by the B\"uchi mechanism,
and are faulty in the sense that counters may erroneously increase at any time.

\begin{narrowfig}{.66\textwidth}
\setlength{\unitlength}{1em}
\begin{picture}(28,6.5)
\node[Nadjust=wh,Nframe=n](LTLF)(6,5.5)
{LTL$^\downarrow_1(\nextt, \sometimes)$}
\node[Nadjust=wh,Nframe=n](LTLU)(23,5.5)
{LTL$^\downarrow_1(\nextt, \until)$}
\node[Nadjust=wh,Nframe=n](CA)(6,1)
{\begin{tabular}{c}
 counter automata \\
 with incrementing errors
\end{tabular}}
\node[Nadjust=wh,Nframe=n](RA)(23,1)
{\begin{tabular}{c}
 one-way alternating \\
 automata with $1$ register
\end{tabular}}
\drawedge(LTLF,LTLU)
{subfragment}
\drawedge[ELside=r,ELpos=45](LTLU,RA)
{log.\ space}
\drawedge[ELside=r](RA,CA)
{\begin{tabular}{c}
 poly.\\
 space
 \end{tabular}}
\drawedge[ELpos=55](CA,LTLF)
{log.\ space}
\end{picture}
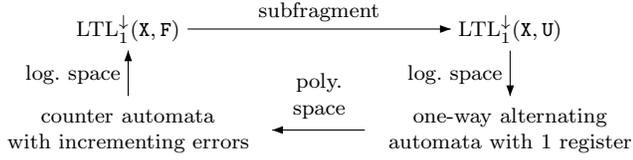
\caption{A circle of translations which preserve language nonemptiness}
\label{f:circle}
\end{narrowfig}

Using results in
\cite{Schnoebelen02,Mayr03,Ouaknine&Worrell06a,Ouaknine&Worrell06c},
we show that nonemptiness for the faulty counter automata is
decidable and not primitive recursive over finite words,
and $\Pi^0_1$-complete over infinite words.
Hence, the same bounds hold for
LTL$^\downarrow_1(\nextt, \sometimes)$ satisfiability,
LTL$^\downarrow_1(\nextt, \until)$ satisfiability, and
nonemptiness of one-way alternating automata with $1$ register.
The latter therefore provide an attractive notion of
regular languages of finite data words,
although the complexity of nonemptiness is very high.

The incrementing errors of counter automata
correspond to restricted powers of
future-time LTL and one-way alternating automata with $1$ register.
As soon as any of $1$ more register,
the $\pastsometimes$ temporal operator or
backward automaton moves are added,
even after replacing $\until$ by $\sometimes$
and restricting to universal automata,
decidability and $\Pi^0_1$-membership break down:
we obtain logarithmic-space translations
from Minsky (i.e., error-free) counter automata,
which result in $\Sigma^0_1$-hardness over finite data words
and $\Sigma^1_1$-hardness over infinite data words.
Together with the bounds via the faulty counter automata,
that gives us a complete complexity table for
fragments of LTL with the freeze quantifier defined by
varying the set of temporal operators and the number of registers
(see Figure~\ref{f:compl.sat}).

Interestingly, similar results were reported in \cite{Ouaknine&Worrell06a}
for real-time metric temporal logic, by translations to and from
alternating automata with $1$ clock and machines with fifo channels.
Indeed, computing a counter automaton with incrementing errors
from a sentence of LTL$^\downarrow_1(\nextt, \until)$
follows the same broad steps as may be employed to
compute a channel machine with insertion errors
from a sentence of (future-time) MTL,
some of which were implicit already in the proof that
whether the language of a one-way nondeterministic register automaton
is contained in the language of such an automaton with $1$ register
is decidable over finite data words \cite[Appendix~A]{Kaminski&Francez94}.
However, there is no obvious translation
from LTL$^\downarrow_1(\nextt, \until)$
to MTL or alternating automata with $1$ clock,
and counters are less powerful than fifo channels.
Also, the translations from counter automata to LTL with freeze in this paper
and those from channel machines to MTL differ substantially.

The decidability of satisfiability for LTL$^\downarrow_1(\nextt, \until)$
over finite data words makes it the competitor of FO$^2(\sim, <, +1)$.
To clarify the relationship between the two logics,
we extend the equiexpressiveness result in \cite{Etessami&Vardi&Wilke02}
and show that FO$^2(\sim, <, +1, \ldots, +m)$ is as expressive as
LTL$^\downarrow_1(\nextt, \nextt^{-1}, \sometimes, \pastsometimes)$
with a restriction on freeze quantification.
However, temporal sentences may be exponentially longer
than equivalent first-order formulae.

\subsubsection*{Organisation}

After setting up machinery in Section~\ref{section-preliminaries}
and presenting the equiexpressiveness result for FO$^2(\sim, <, +1, \ldots, +m)$
in Section~\ref{s:LTL.FO}, the core of the paper is
Sections \ref{section-upper-bounds} and \ref{section-lower-bounds},
which contain the results on complexity of
satisfiability for fragments of LTL with the freeze quantifier and
nonemptiness for classes of register automata.
We conclude in Section~\ref{section-conclusion}.

\section{Preliminaries}
\label{section-preliminaries}

After defining words with data,
we introduce below logics and automata with which we shall work in the paper.
To define acceptance by alternating automata,
we recall a simple class of two-player games.
This section also contains several results which will be used later.
Their proofs are either relatively straightforward or
heavily based on proofs in the literature.

\subsection{Data Words}

A \emph{data word} $\adataword$ over a finite alphabet $\aalphabet$ is
a nonempty word $\mathrm{str}(\adataword)$ over $\aalphabet$ together with
an equivalence relation $\sim^\adataword$ on its positions.
We write $\length{\adataword}$ for the length of the word,
$\adataword(i)$ for the letter at position $i$, and
$[i]_{\sim^\adataword}$ for the class that contains $i$,
where $0 \leq i < \length{\adataword}$.
When $\adataword$ is understood, we may write
simply $\sim$ instead of $\sim^\adataword$.
We shall sometimes refer to classes of $\sim$ as `data'.

\begin{example}
\label{ex:data.word}
A data word of length $3$ over $\{\aletter, \aletterbis\}$ is $\adataword$
such that $\mathrm{str}(\adataword) = \aletter \aletter \aletterbis$
and the classes of $\sim^\adataword$ are $\{0, 2\}$ and $\{1\}$.
\end{example}

\subsection{LTL over Data Words}
%\label{section-LTL} 

\subsubsection*{Syntax}

LTL$^\downarrow(\mathcal{O})$
will denote the linear temporal logic with the freeze quantifier
and temporal operators in the set $\mathcal{O}$.
Each formula is over a finite alphabet $\aalphabet$.
Atomic propositions $\aletter$ are elements of $\aalphabet$,
$\atempop$ ranges over $\mathcal{O}$, and
$r$ ranges over $\mathbb{N}_{> 0}$.
\[
\aformula \:::=\:
\aletter \,\mid\, \top \,\mid\,
\neg \aformula \,\mid\, \aformula \wedge \aformula \,\mid\,
\atempop(\aformula, \ldots, \aformula) \,\mid\,
{\downarrow_r} \aformula \,\mid\, {\uparrow_r}
\]

An occurence of $\uparrow_r$ within the scope of
some freeze quantification $\downarrow_r$ is bound by it;
otherwise, it is free.
A sentence is a formula with no free occurence of any $\uparrow_r$.

We consider temporal operators
`next' ($\nextt$), `eventually' ($\sometimes$), `until' ($\until$),
and their past-time versions ($\nextt^{-1}, \pastsometimes, \since$).
As $\sometimes \aformula$ is equivalent to $\top \until \aformula$,
$\sometimes$ can be omitted from any set which contains $\until$,
and the same is true for $\pastsometimes$ and $\since$.
As usual, we regard $\always$ (`always') and $\pastalways$ (`past always')
as abbreviations for $\neg \sometimes \neg$ and $\neg \pastsometimes \neg$.

Let LTL$^\downarrow_n(\mathcal{O})$ be the fragment with $n$ registers,
i.e.\ where $r \in \{1, \ldots, n\}$.

\subsubsection*{Semantics}

A \emph{register valuation} $\aregval$ for a data word $\adataword$ is
a finite partial map from $\mathbb{N}_{> 0}$
to the classes in $\adataword$,
i.e.\ to $\{i \,:\, 0 \leq i < \length{\adataword}\} / {\sim}$.
If $r \notin \mathrm{dom}(\aregval)$, then the atomic formula $\uparrow_r$
will evaluate to false with respect to $\aregval$.
Such undefined register values will be used for initial automata states.
We say that $\aregval$ is an $n$-register valuation iff
$\mathrm{dom}(\aregval) \subseteq \{1, \ldots, n\}$.

For a data word $\adataword$ over a finite alphabet $\aalphabet$,
a position $0 \leq i < \length{\adataword}$,
a register valuation $\aregval$ for $\adataword$, and
a formula $\aformula$ over $\aalphabet$,
writing $\adataword, i \,\models_\aregval\, \aformula$ will mean that
$\aformula$ is satisfied by $\adataword$ at position $i$
with respect to $\aregval$.
The satisfaction relation is defined as follows,
where we omit the Boolean cases.
\begin{eqnarray*}
\adataword, i \,\models_\aregval\, \aletter
& \equivdef &
\adataword(i) = \aletter
\\
\adataword, i \,\models_\aregval\, \nextt \aformula
& \equivdef &
i + 1 < \length{\adataword}\ \mathrm{and}\ 
\adataword, i + 1 \,\models_\aregval\, \aformula
\\
\adataword, i \,\models_\aregval\, \nextt^{-1} \aformula
& \equivdef &
i - 1 \geq 0\ \mathrm{and}\ 
\adataword, i - 1 \,\models_\aregval\, \aformula
\\
\adataword, i \,\models_\aregval\, \aformula \until \aformulabis
& \equivdef &
\mathrm{for\ some}\ j \geq i,\ 
\adataword, j \,\models_\aregval\, \aformulabis\ \mathrm{and}\ 
\mathrm{for\ all}\ i \leq j' < j,\ 
\adataword, j' \,\models_\aregval\, \aformula
\\
\adataword, i \,\models_\aregval\, \aformula \since \aformulabis
& \equivdef &
\mathrm{for\ some}\ j \leq i,\ 
\adataword, j \,\models_\aregval\, \aformulabis\ \mathrm{and}\ 
\mathrm{for\ all}\ j < j' \leq i,\ 
\adataword, j' \,\models_\aregval\, \aformula
\\
\adataword, i \,\models_\aregval\, {\downarrow_r} \aformula
& \equivdef &
\adataword, i \,\models_{\aregval[r \mapsto [i]_\sim]}\, \aformula
\\
\adataword, i \,\models_\aregval\, {\uparrow_r}
& \equivdef &
r \in \mathrm{dom}(\aregval)\ \mathrm{and}\ 
i \in \aregval(r)
\end{eqnarray*}

\begin{example}
\label{ex:LTL}
Consider the sentence
$\aformula =
 \always \big(\aletter \Rightarrow {\downarrow_1} \nextt
              \big((\always (\aletter \Rightarrow \neg {\uparrow_1})) \wedge
                   (\sometimes (\aletterbis \wedge {\uparrow_1}))\big)\big)$
of LTL$^\downarrow_1(\nextt, \sometimes)$,
which is over the alphabet $\{\aletter, \aletterbis\}$.
It states that no two letters $\aletter$ are in the same class,
and that every letter $\aletter$ is followed by
a letter $\aletterbis$ which is in the same class.
Thus, for the data word $\adataword$ in Example~\ref{ex:data.word},
we have $\adataword, 0 \,\not\models_\emptyset\, \aformula$.
\end{example}

\subsection{FO over Data Words}

As defined in \cite{Bojanczyketal06a},
FO$(\sim, <, +1, \ldots, +m)$ denotes first-order logic over data words,
in which variables range over word positions.
We use variable names $x_0$, $x_1$, \ldots \
The predicates $x_i < x_j$ and $x_i = x_j + k$ are interpreted as expected.
Each formula has an alphabet $\aalphabet$, and it may contain
unary predicates $P_\aletter(x_i)$ which are satisfied by a data word iff
the letter at position $x_i$ is $\aletter$.
When we write $\aformula(x_{i_1}, \ldots, x_{i_N})$, it means that
at most $x_{i_1}$, \ldots, $x_{i_N}$ occur free in $\aformula$.

FO$^n(\sim, <, +1, \ldots, +m)$ is the fragment with $n$ variables
$x_0$, \ldots, $x_{n - 1}$.

\begin{example}
\label{ex:FO}
Let $\aformula'(x_0)$ be the following formula of FO$^2(\sim, <)$,
which states that, from position $x_0$ onwards,
no two letters $\aletter$ are in the same class,
and every letter $\aletter$ is followed by
a letter $\aletterbis$ which is in the same class.
\[\begin{array}{c}
\fforall{x_1}{\neg (x_1 < x_0) \wedge P_\aletter(x_1) \,\Rightarrow\\
              \fforall{x_0}{x_1 < x_0 \wedge P_\aletter(x_0) \,\Rightarrow\,
                            \neg x_1 \sim x_0} \wedge
              \eexists{x_0}{x_1 < x_0 \wedge P_\aletterbis(x_0) \wedge
                            x_1 \sim x_0}}
\end{array}\]
It is equivalent to the sentence $\aformula$ from Example~\ref{ex:LTL}
in the sense that,
for every data word $\adataword$ over $\{\aletter, \aletterbis\}$
and $0 \leq i < \length{\adataword}$,
we have
$\adataword, i \,\models_\emptyset\, \aformula$ iff
$\adataword \models_{[x_0 \mapsto i]} \aformula'(x_0)$.
\end{example}

\subsection{Weak Games}

The automata that will be introduced in the next section will be
alternating and weak \cite{Muller&Saoudi&Schupp86},
so we shall use the following class of
zero-sum two-player finitely branching games
to define acceptance by such automata.

\subsubsection*{Games}

A \emph{weak game} $\agame$ is a tuple
$\tuple{\poss, \poss_1, \poss_2, \rightarrow, \rank}$ such that:
\begin{itemize}
\item
$\poss$ is a set of all positions;
\item
$\poss_1$ and $\poss_2$ disjointly partition $\poss$ into
positions of players $1$ and $2$ (respectively);
\item
${\rightarrow} \,\subseteq\, \poss \times \poss$ is a successor relation
with respect to which every position has finitely many successors;
\item
$\rank: \poss \rightarrow \mathbb{N}$ specifies ranks so that,
whenever $\apos \rightarrow \apos'$,
we have $\rank(\apos) \geq \rank(\apos')$.
\end{itemize}

A play $\aplay$ of $\agame$ is
a sequence $\apos_0 \apos_1 \ldots$ of positions of $\agame$
such that $\apos_i \rightarrow \apos_{i + 1}$ for each $i$.
If $\aplay$ is infinite, let $\rank(\aplay) = \rank(\apos_i)$,
where $i$ is such that $\rank(\apos_j) = \rank(\apos_i)$ for all $j > i$
(such an $i$ necessarily exists).

We say that a play $\aplay$ of $\agame$ is complete iff
either it ends with a position without successors
or it is infinite.
For such $\aplay$, we consider it winning for player $1$ iff
either it ends with a position of player $2$
or it is infinite and $\rank(\aplay)$ is even.
The winning condition for player $2$ is symmetric,
with the opposite parity.

A strategy for player $l$ from a position $\apos$ of $\agame$ is
a tree $\astr \subseteq \poss^{< \omega}$ of finite plays of $\agame$ such that:
\begin{itemize}
\item[(i)]
$\apos \in \astr$ and it is the root;
\item[(ii)]
whenever $\aplay \in \astr$ ends with a position $\apos$ of player $l$
which has at least one successor, it has a unique child;
\item[(iii)]
whenever $\aplay \in \astr$ ends with a position $\apos$ of the other player,
it has all children $\aplay \apos'$ with $\apos \rightarrow \apos'$.
\end{itemize}
We say that $\astr$ is positional iff the choices of successors in (ii)
depend only on the ending positions $\apos$.

Now, a play by $\astr$ is either an element of $\astr$
or an infinite sequence whose every nonempty prefix is an element of $\astr$.
We say that $\astr$ is winning iff
each complete play by $\astr$ is winning for player $l$.

\subsubsection*{Consistent Signature Assignments}

Let $\agame$ be a weak game as above.

A consistent signature assignment for $\agame$ is
a function $\ass$ from some $\winn \subseteq \poss$ to $\mathbb{N}$
such that the following are satisfied,
where pairs of natural numbers are ordered lexicographically,
i.e.\ $\tuple{n, m} < \tuple{n', m'}$ iff
either $n < n'$, or $n = n'$ and $m < m'$.
\begin{itemize}
\item
for every $\apos \,\in\, \winn \cap \poss_1$,
there exists $\apos \rightarrow \apos'$ with
$\apos' \in \winn$ and
$\tuple{\rank(\apos'), \ass(\apos')} \leq \tuple{\rank(\apos), \ass(\apos)}$,
where the inequality is strict if $\rank(\apos)$ is odd;
\item
for every $\apos \,\in\, \winn \cap \poss_2$
and every $\apos \rightarrow \apos'$, we have
$\apos' \in \winn$ and
$\tuple{\rank(\apos'), \ass(\apos')} \leq \tuple{\rank(\apos), \ass(\apos)}$,
where the inequality is strict if $\rank(\apos)$ is odd.
\end{itemize}

Part~(a) of the result below is straightforward,
whereas part~(b) is obtained by simplifying the proof of
\cite[Lemma~10]{Walukiewicz01} which is for more general parity games.

\begin{theorem}
\label{th:WG}
Suppose $\agame$ is a weak game.
\begin{itemize}
\item[(a)]
For every consistent signature assignment $\ass$ for $\agame$
and every $\apos \in \mathrm{dom}(\ass)$,
player $1$ has a positional winning strategy from $\apos$.
\item[(b)]
There exists a consistent signature assignment $\ass$ for $\agame$
such that for every $\apos \notin \mathrm{dom}(\ass)$,
player $2$ has a positional winning strategy from $\apos$.
\end{itemize}
\end{theorem}

The following are two immediate corollaries:
\begin{itemize}
\item
positional determinacy,
i.e.\ that for every position $\apos$,
one of the players has a positional winning strategy from $\apos$;
\item
for every position $\apos$,
there exists a consistent signature assignment which is defined for $\apos$ iff
player $1$ has a positional winning strategy from $\apos$.
\end{itemize}

\subsection{Register Automata}
%\label{section-regaut}

Corresponding to the addition of the freeze quantifier to LTL,
finite automata can be extended by registers.
We now define two-way alternating register automata over data words.

A state of such an automaton for a data word will consist of
a word position, an automaton location and a register valuation.
From it, according to the transition function,
one of the following is performed:
\begin{itemize}
\item
branching to another location depending on one of the following Boolean tests:
whether the current letter equals a specified letter,
whether the word position is the first or last, or
whether the current datum equals the datum in a specified register;
\item
storing the current datum into a register;
\item
conjunctive or disjunctive branching to a pair of locations;
\item
acceptance or rejection;
\item
moving to the next or previous word position.
\end{itemize}

The automata will be weak in that each location will have a rank,
which will not increase after any transition,
and whose parities will be used to define acceptance.

Each location will also have a height, which will decrease after
every transition which is not a move to another word position.
The heights ensure that infinite progress cannot be made
while remaining at the same word position.
That constraint simplifies some proofs without reducing expressiveness.

\begin{remark}
\label{rem:RA}
In contrast to the formalisations of register automata in
\cite{Kaminski&Francez94,Sakamoto&Ikeda00,Neven&Schwentick&Vianu04},
data stored in registers within an automaton state will not be required
to be mutually distinct and to contain the datum from
the previously visited word position.%
\footnote{That is a minor technical difference.  It can be checked that,
for every automaton with $n + 1$ registers in the sense of
\cite{Kaminski&Francez94,Sakamoto&Ikeda00,Neven&Schwentick&Vianu04},
one can construct an equivalent automaton with $n + 1$ registers and
an equivalent alternating automaton with $n$ registers
in the sense of this paper.}
\end{remark}

\subsubsection*{Automata}

The set $\Delta(\aalphabet, \locs, n)$ of all transition formulae
over a finite alphabet $\aalphabet$,
over a finite set $\locs$ of locations and
with $n \in \mathbb{N}$ registers
is defined below.
\begin{eqnarray*}
B(\aalphabet, n) & = &
\{\aletter, \mathtt{beg}, \mathtt{end}, \uparrow_r
  \,:\,
  \aletter \in \aalphabet,
  r \in \{1, \ldots, n\}\}
\\
\Delta(\aalphabet, \locs, n) & = &
\{\ifte{\beta}{\aloc}{\aloc'},
  {\downarrow_r} \aloc,
  \aloc \wedge \aloc', \aloc \vee \aloc',
  \top, \bot,
  \nextt \aloc, \overline{\nextt} \aloc,
  \nextt^{-1} \aloc, \overline{\nextt^{-1}} \aloc
  \,:\\ & &
  \beta \in B(\aalphabet, n),
  \aloc, \aloc' \in \locs,
  r \in \{1, \ldots, n\}\}
\end{eqnarray*}
We have that $\Delta(\aalphabet, \locs, n)$ is closed under
the self-inverse operation of taking duals:
\[\begin{array}{rcl@{\hspace{2em}}rcl@{\hspace{2em}}rcl}
\overline{\ifte{\beta}{\aloc}{\aloc'}} & = &
\ifte{\beta}{\aloc}{\aloc'}
&
\overline{\aloc \wedge \aloc'} & = &
\aloc \vee \aloc'
&
\overline{\nextt \aloc} & = &
\overline{\nextt} \aloc
\\
\overline{{\downarrow_r} \aloc} & = &
{\downarrow_r} \aloc
&
\overline{\top} & = &
\bot
&
\overline{\nextt^{-1} \aloc} & = &
\overline{\nextt^{-1}} \aloc
\end{array}\]
The difference between transition formulae
$\nextt \aloc$ and their duals $\overline{\nextt} \aloc$
is that the former will be rejecting and the latter accepting
if there is no next word position,
and similarly for $\nextt^{-1} \aloc$ and $\overline{\nextt^{-1}} \aloc$.

A \emph{register automaton} $\aregaut$ is a tuple
$\tuple{\aalphabet, \locs, \aloc_I, n, \delta, \rank, \height}$ as follows:
\begin{itemize}
\item
$\aalphabet$ is a finite alphabet;
\item
$\locs$ is a finite set of locations;
\item
$\aloc_I \in \locs$ is the initial location;
\item
$n \in \mathbb{N}$ is the number of registers;
\item
$\delta: \locs \rightarrow \Delta(\aalphabet, \locs, n)$
is a transition function;
\item
$\rank: \locs \rightarrow \mathbb{N}$ specifies ranks and is such that,
whenever $\aloc'$ occurs in $\delta(\aloc)$,
we have $\rank(\aloc') \leq \rank(\aloc)$;
\item
$\height: \locs \rightarrow \mathbb{N}$ specifies heights and is such that,
whenever $\delta(\aloc)$ is of the form
$\ifte{\beta}{\aloc'}{\aloc''}$,
${\downarrow_r} \aloc'$,
$\aloc' \wedge \aloc''$ or $\aloc' \vee \aloc''$,
we have $\height(\aloc'), \height(\aloc'') < \height(\aloc)$.
\end{itemize}

We say that a register automaton is:
\begin{describe}{\emph{nondeterministic}}
\item[\hfill\emph{one-way}]
iff no $\delta(\aloc)$ is of the form
$\ifte{\mathtt{beg}}{\aloc'}{\aloc''}$,
$\nextt^{-1} \aloc'$ or $\overline{\nextt^{-1}} \aloc'$;
\item[\emph{nondeterministic}]
iff no $\delta(\aloc)$ is of the form
$\aloc' \wedge \aloc''$;
\item[\hfill\emph{universal}]
iff no $\delta(\aloc)$ is of the form
$\aloc' \vee \aloc''$;
\item[\hfill\emph{deterministic}]
iff it is both nondeterministic and universal.
\end{describe}
For $d \in \{\mbox{1}, \mbox{2}\}$
and $C \in \{\mbox{A}, \mbox{N}, \mbox{U}, \mbox{D}\}$,
let $dC$RA denote the class of all register automata
with restrictions on directionality and control specified by $d$ and $C$.
Let $dC$RA$_n$ denote the subclass with $n$ registers.

\subsubsection*{Acceptance Games}

Let $\aregaut$ be a register automaton as above,
and $\adataword$ be a data word over $\aalphabet$.
The acceptance game of $\aregaut$ over $\adataword$
is the weak game
$\agame_{\aregaut, \adataword} =
 \tuple{\poss, \poss_1, \poss_2, \rightarrow, \rank}$
defined below.
Player $1$ (``automaton'') will be
resolving the disjunctive branchings
given by the transition function of $\aregaut$,
winning a finite play if it ends with an accepting state, and
winning an infinite play if the limit location rank is even.
Dually, player $2$ (``pathfinder'') will be
resolving the conjunctive branchings and
winning at rejecting states or by odd limit ranks.
\begin{itemize}
\item
$\poss$ is the set of all states of $\aregaut$ for $\adataword$,
which are triples $\tuple{i, \aloc, \aregval}$ where
$0 \leq i < \length{\adataword}$, $\aloc \in \locs$, and
$\aregval$ is an $n$-register valuation for $\adataword$.
\item
The partition of $\poss$ into $\poss_1$ and $\poss_2$,
and the successor relation, are given by the table in Figure~\ref{f:acc.games}.
The ownership of states with unique successors
has not been specified because it is irrelevant.
The table omits dual transition formulae,
which are treated by swapping the ownerships.
\item
For every $\tuple{i, \aloc, \aregval} \in \poss$,
$\rank(\tuple{i, \aloc, \aregval}) = \rank(\aloc)$.
\end{itemize}
Observe that every branching in $\agame_{\aregaut, \adataword}$
is at most binary.

\begin{figure}
\[\begin{array}{r|c|c|}
\delta(\aloc) &
\mathrm{owner\ of\ } \tuple{i, \aloc, \aregval} &
\mathrm{successors\ of\ } \tuple{i, \aloc, \aregval}
\\ \hline
\ifte{\beta}{\aloc'}{\aloc''} &
&
\{\tuple{i, \aloc', \aregval}\},
\mathrm{\ if\ }
\adataword, i \,\models_\aregval\, \beta
\hspace{2em}
\{\tuple{i, \aloc'', \aregval}\},
\mathrm{\ if\ }
\adataword, i \,\not\models_\aregval\, \beta
\\ \hline
{\downarrow_r} \aloc' &
&
\{\tuple{i, \aloc', \aregval[r \mapsto [i]_\sim]}\}
\\ \hline
\aloc' \wedge \aloc'' &
2 &
\{\tuple{i, \aloc', \aregval}, \tuple{i, \aloc'', \aregval}\}
\\ \hline
\top &
2 &
\emptyset
\\ \hline
\nextt \aloc' &
1, \mathrm{\ if\ } i + 1 = \length{\adataword} &
\{\tuple{i + 1, \aloc', \aregval}\}, \mathrm{\ if\ } i + 1 < \length{\adataword}
\hspace{2em}
\emptyset, \mathrm{\ if\ } i + 1 = \length{\adataword}
\\ \hline
\nextt^{-1} \aloc' &
1, \mathrm{\ if\ } i = 0 &
\{\tuple{i - 1, \aloc', \aregval}\}, \mathrm{\ if\ } i > 0
\hspace{2em}
\emptyset, \mathrm{\ if\ } i = 0
\\ \hline
\end{array}\]
\[\begin{array}{rcl@{\hspace{2em}}rcl}
\adataword, i \,\models_\aregval\, \mathtt{beg}
& \equivdef &
i = 0
&
\adataword, i \,\models_\aregval\, \mathtt{end}
& \equivdef &
i + 1 = \length{\adataword}
\end{array}\]
\caption{Defining acceptance games}
\label{f:acc.games}
\end{figure}

A run of $\aregaut$ over $\adataword$ is
a strategy $\astr$ in $\agame_{\aregaut, \adataword}$ for player $1$
from the initial state $\tuple{0, \aloc_I, \emptyset}$.
We say that $\astr$ is accepting iff it is winning,
and that $\aregaut$ accepts $\adataword$ iff
$\aregaut$ has an accepting run over $\adataword$.

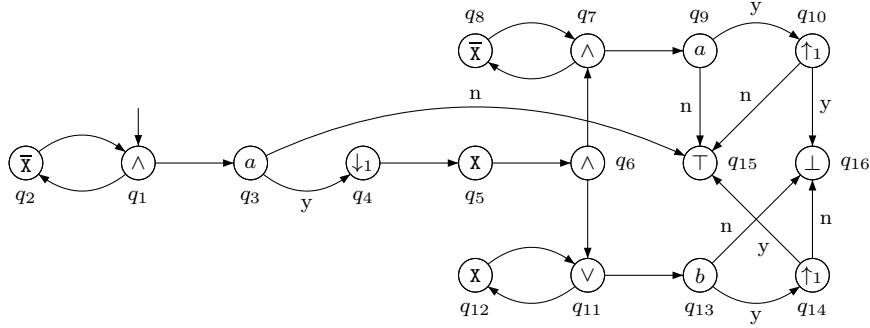
\begin{figure}
\setlength{\unitlength}{2.5em}
\begin{center}
\begin{picture}(16,5.5)(0,.25)
\gasset{Nw=.6,Nh=.6,Nmr=.3,ExtNL=y,NLangle=270,NLdist=.2}
\node[Nmarks=i,iangle=90](1)(3,3){$\aloc_1$}
\node(2)(1,3){$\aloc_2$}
\node(3)(5,3){$\aloc_3$}
\node(4)(7,3){$\aloc_4$}
\node(5)(9,3){$\aloc_5$}
\node[NLangle=0](6)(11,3){$\aloc_6$}
\node[NLangle=90](7)(11,5){$\aloc_7$}
\node[NLangle=90](8)(9,5){$\aloc_8$}
\node[NLangle=90](9)(13,5){$\aloc_9$}
\node[NLangle=90](10)(15,5){$\aloc_{10}$}
\node(11)(11,1){$\aloc_{11}$}
\node(12)(9,1){$\aloc_{12}$}
\node(13)(13,1){$\aloc_{13}$}
\node(14)(15,1){$\aloc_{14}$}
\node[NLangle=0](15)(13,3){$\aloc_{15}$}
\node[NLangle=0](16)(15,3){$\aloc_{16}$}
\node[ExtNL=n,NLdist=0](1)(3,3){$\wedge$}
\node[ExtNL=n,NLdist=0](2)(1,3){$\overline{\nextt}$}
\node[ExtNL=n,NLdist=0](3)(5,3){$\aletter$}
\node[ExtNL=n,NLdist=0](4)(7,3){$\downarrow_1$}
\node[ExtNL=n,NLdist=0](5)(9,3){$\nextt$}
\node[ExtNL=n,NLdist=0](6)(11,3){$\wedge$}
\node[ExtNL=n,NLdist=0](7)(11,5){$\wedge$}
\node[ExtNL=n,NLdist=0](8)(9,5){$\overline{\nextt}$}
\node[ExtNL=n,NLdist=0](9)(13,5){$\aletter$}
\node[ExtNL=n,NLdist=0](10)(15,5){$\uparrow_1$}
\node[ExtNL=n,NLdist=0](11)(11,1){$\vee$}
\node[ExtNL=n,NLdist=0](12)(9,1){$\nextt$}
\node[ExtNL=n,NLdist=0](13)(13,1){$\aletterbis$}
\node[ExtNL=n,NLdist=0](14)(15,1){$\uparrow_1$}
\node[ExtNL=n,NLdist=0](15)(13,3){$\top$}
\node[ExtNL=n,NLdist=0](16)(15,3){$\bot$}
\drawedge[curvedepth=.5](1,2){}
\drawedge[curvedepth=.5](2,1){}
\drawedge(1,3){}
\drawedge[curvedepth=-.5,ELside=r](3,4){y}
\drawedge[curvedepth=1](3,15){n}
\drawedge(4,5){}
\drawedge(5,6){}
\drawedge(6,7){}
\drawedge[curvedepth=.5](7,8){}
\drawedge[curvedepth=.5](8,7){}
\drawedge(7,9){}
\drawedge[curvedepth=.5](9,10){y}
\drawedge[ELside=r](9,15){n}
\drawedge(10,16){y}
\drawedge[ELside=r](10,15){n}
\drawedge(6,11){}
\drawedge[curvedepth=.5](11,12){}
\drawedge[curvedepth=.5](12,11){}
\drawedge(11,13){}
\drawedge[curvedepth=-.5,ELside=r](13,14){y}
\drawedge[ELpos=33](13,16){n}
\drawedge[ELpos=33](14,15){y}
\drawedge[ELside=r](14,16){n}
\end{picture}
\end{center}
\caption{A register automaton}
\label{f:RA}
\end{figure}

\begin{figure}
\setlength{\unitlength}{3.5em}
\begin{center}
\begin{picture}(12,3)(0,-5.5)
\gasset{Nw=.6,Nh=.6,Nmr=.3,Nadjust=wh,ExtNL=n,NLdist=0}
\node(01e)(3,-5){$0, \aloc_1, -$}
\node[Nframe=n](02e)(1,-4){$0, \aloc_2, -$}
\node(11e)(3,-3){$1, \aloc_1, -$}
\node[Nframe=n](13e)(5,-3){$1, \aloc_3, -$}
\node[Nframe=n](14e)(7,-3){$1, \aloc_4, -$}
\node[Nframe=n](151)(9,-3){$1, \aloc_5, \{1\}$}
\node(261)(11,-3){$2, \aloc_6, \{1\}$}
\node[Nmr=0](2111)(7,-5){$2, \aloc_{11}, \{1\}$}
\node[Nmr=0](2121)(5,-5){$2, \aloc_{12}, \{1\}$}
\node[Nframe=n](2131)(9,-5){$2, \aloc_{13}, \{1\}$}
\node[Nframe=n](2141)(11,-5){$2, \aloc_{14}, \{1\}$}
\node[Nmr=0](2161)(11,-4){$2, \aloc_{16}, \{1\}$}
\drawedge[curvedepth=.5](01e,02e){}
\drawedge[curvedepth=.5](02e,11e){}
\drawedge(11e,13e){}
\drawedge[curvedepth=-.5](13e,14e){}
\drawedge(14e,151){}
\drawedge(151,261){}
\drawedge(261,2111){}
\drawedge[curvedepth=.5](2111,2121){}
\drawedge(2111,2131){}
\drawedge[curvedepth=-.5](2131,2141){}
\drawedge(2141,2161){}
\end{picture}
\end{center}
\caption{A positional winning strategy}
\label{f:pos.win.str}
\end{figure}
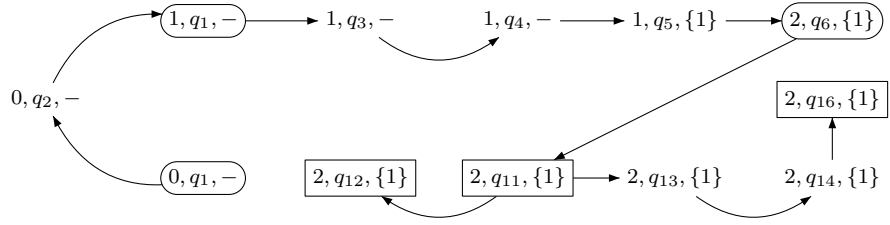

\begin{example}
\label{ex:RA}
Let $\aregaut$ be a register automaton
with alphabet $\{\aletter, \aletterbis\}$ and $1$ register,
whose locations and transition function are shown in Figure~\ref{f:RA},
and such that the ranks of $\aloc_1$ and $\aloc_7$ are even
but the rank of $\aloc_{11}$ is odd.
It is straightforward to assign exact ranks and heights
to the locations of $\aregaut$ so that the conditions
in the definition of register automata are satisfied.

We have that $\aregaut$ is one-way, neither nondeterministic nor universal,
and equivalent to the sentence $\aformula$ from Example~\ref{ex:LTL}
in the sense that, for every data word
$\adataword$ over $\{\aletter, \aletterbis\}$,
$\aregaut$ accepts $\adataword$ iff
$\adataword, 0 \,\models_\emptyset\, \aformula$.

In particular, $\aregaut$ rejects
the data word $\adataword$ from Example~\ref{ex:data.word}.
By positional determinacy (cf.\ Theorem~\ref{th:WG})
of the acceptance game $\agame_{\aregaut, \adataword}$,
player $2$ (``pathfinder'') has a positional winning strategy
from the initial state $\tuple{0, \aloc_1, \emptyset}$.
Such a strategy is shown in Figure~\ref{f:pos.win.str},
where $-$ and $\{1\}$ abbreviate register valuations
$\emptyset$ and $[1 \mapsto \{1\}]$ (respectively),
sharp and oval frames indicate states belonging to
players $1$ and $2$ (respectively),
and states whose owner is irrelevant are not framed.
The strategy is positional trivially,
as no state is visited more than once.
Essentially, the pathfinder challenges the automaton
to find a letter $\aletterbis$ which follows the second letter $\aletter$
and is in the same class.
\end{example}

\subsubsection*{Closure Properties}

We now consider closure of classes of register automata
under complement, intersection and union.

\begin{theorem}
\label{th:closure}
\begin{itemize}
\item[(a)]
For each $d \in \{\mbox{1}, \mbox{2}\}$,
$d$ARA$_n$ and $d$DRA$_n$ are closed under complement,
and $d$NRA$_n$ is dual to $d$URA$_n$.
\item[(b)]
For each $C \in \{\mbox{A}, \mbox{N}, \mbox{U}, \mbox{D}\}$,
1$C$RA is closed under intersection and union.
For intersections of universal or alternating automata, and
for unions of nondeterministic or alternating automata,
the maximum of the two numbers of registers suffices.
Otherwise, their sum suffices.
\item[(c)]
2URA is closed under intersection,
2NRA is closed under union, and
2ARA is closed under intersection and union.
The maximum of the two numbers of registers suffices.
\end{itemize}
\noindent
In each case, a required automaton is computable in logarithmic space.
\end{theorem}

\begin{proof}
For a register automaton $\aregaut$ as above, its dual
$\overline{\aregaut} =
 \tuple{\aalphabet, \locs, \aloc_I, n,
        \overline{\delta}, \overline{\rank}, \height}$
is defined by
$\overline{\delta}(\aloc) = \overline{\delta(\aloc)}$ and
$\overline{\rank}(\aloc) = \rank(\aloc) + 1$
for each $\aloc \in \locs$.

It suffices for (a) to show that,
for every data word $\adataword$ over $\aalphabet$,
$\aregaut$ accepts $\adataword$ iff
$\overline{\aregaut}$ rejects $\adataword$.
The latter is immediate by determinacy of weak games
(cf.~Theorem~\ref{th:WG}), and by observing that
$\astr$ is a winning strategy in $\agame_{\aregaut, \adataword}$
for player $1$ from $\tuple{0, \aloc_I, \emptyset}$ iff
$\astr$ is a winning strategy in $\agame_{\overline{\aregaut}, \adataword}$
for player $2$ from $\tuple{0, \aloc_I, \emptyset}$.

The nontrivial parts of (b) and (c) are the closures of
1NRA under intersection, 1URA under union,
and 1DRA under intersection and union.
By (a), we shall be done if, given
$\aregaut_1 =
 \tuple{\aalphabet, \locs_1, \aloc^1_I, n_1, \delta_1, \rank_1, \height_1}$ and
$\aregaut_2 =
 \tuple{\aalphabet, \locs_2, \aloc^2_I, n_2, \delta_2, \rank_2, \height_2}$
in 1NRA, we show how to compute in logarithmic space
$\aregaut =
 \tuple{\aalphabet, \locs, \aloc_I, n_1 + n_2, \delta, \rank, \height}$
in 1NRA which accepts a data word $\adataword$ over $\aalphabet$ iff
both $\aregaut_1$ and $\aregaut_2$ do,
and such that $\aregaut$ is in 1DRA if both $\aregaut_1$ and $\aregaut_2$ are.

As in the proof of \cite[Theorem~3]{Kaminski&Francez94},
$\aregaut$ is obtained by a product construction, so we only provide it.
The locations of $\aregaut$ are
pairs of locations of $\aregaut_1$ and $\aregaut_2$,
i.e.\ $\locs = \locs_1 \times \locs_2$,
and the initial location $\aloc_I$ is $\tuple{\aloc^1_I, \aloc^2_I}$.
Transitions of $\aregaut$ will be of one of the following three kinds:
\begin{itemize}
\item
a transition of one of $\aregaut_1$ or $\aregaut_2$
which does not change the word position;
\item
a transition of one of $\aregaut_1$ or $\aregaut_2$
which moves to the next word position,
provided the other automaton has accepted;
\item
a pair of transitions of $\aregaut_1$ and $\aregaut_2$
which both move to the next word position.
\end{itemize}
For the former two kinds, we define $\tuple{\atransf_1, \aloc_2}$
for a transition formula $\atransf_1$ of $\aregaut_1$
and a location $\aloc_2$ of $\aregaut_2$
to be the transition formula of $\aregaut$ obtained by
pairing with $\aloc_2$ each location which occurs in $\atransf_1$.
Transition formulae $\tuple{\aloc_1, \atransf_2}$ are defined similarly,
where also occurences of registers $r_2$ in $\atransf_2$
are replaced by $n_1 + r_2$, so e.g.\
$\tuple{\aloc_1, \ifte{{\uparrow_{r_2}}}{\aloc'_2}{\aloc''_2}} =
 \ifte{{\uparrow_{n_1 + r_2}}}
      {\tuple{\aloc_1, \aloc'_2}}
      {\tuple{\aloc_1, \aloc''_2}}$.
For each $\tuple{\aloc_1, \aloc_2} \in \locs$,
its transition formula is then defined in Figure~\ref{f:int.1NRA},
where the choice in the upper left-hand corner of
$\tuple{\delta_1(\aloc_1), \aloc_2}$ instead of
$\tuple{\aloc_1, \delta_2(\aloc_2)}$ is arbitrary.
The ranks are given by
$\rank(\tuple{\aloc_1, \aloc_2}) =
 (\rank(\aloc_1) + 1) \times (\rank(\aloc_2) + 1) + 1$,
which is even iff $\rank(\aloc_1)$ and $\rank(\aloc_2)$ are both even,
and the heights by
$\height(\tuple{\aloc_1, \aloc_2}) =
 \height(\aloc_1) + \height(\aloc_2)$.
\end{proof}

\begin{figure}
\[\begin{array}{r|c|c|c|c|c|}
&
\ifte{\beta_2}{\aloc'_2}{\aloc''_2},
{\downarrow_{r_2}} \aloc'_2,
\aloc'_2 \vee \aloc''_2 &
\top &
\bot &
\nextt \aloc'_2 &
\overline{\nextt} \aloc'_2
\\ \hline
\begin{array}{r}
\ifte{\beta_1}{\aloc'_1}{\aloc''_1}, \\
{\downarrow_{r_1}} \aloc'_1, \\
\aloc'_1 \vee \aloc''_1
\end{array} &
\tuple{\delta_1(\aloc_1), \aloc_2} &
\tuple{\delta_1(\aloc_1), \aloc_2} &
\bot &
\tuple{\delta_1(\aloc_1), \aloc_2} &
\tuple{\delta_1(\aloc_1), \aloc_2}
\\ \hline
\top &
\tuple{\aloc_1, \delta_2(\aloc_2)} &
\top &
\bot &
\tuple{\aloc_1, \delta_2(\aloc_2)} &
\tuple{\aloc_1, \delta_2(\aloc_2)}
\\ \hline
\bot &
\bot &
\bot &
\bot &
\bot &
\bot
\\ \hline
\nextt \aloc'_1 &
\tuple{\aloc_1, \delta_2(\aloc_2)} &
\tuple{\delta_1(\aloc_1), \aloc_2} &
\bot &
\nextt \tuple{\aloc'_1, \aloc'_2} &
\nextt \tuple{\aloc'_1, \aloc'_2}
\\ \hline
\overline{\nextt} \aloc'_1 &
\tuple{\aloc_1, \delta_2(\aloc_2)} &
\tuple{\delta_1(\aloc_1), \aloc_2} &
\bot &
\nextt \tuple{\aloc'_1, \aloc'_2} &
\overline{\nextt} \tuple{\aloc'_1, \aloc'_2}
\\ \hline
\end{array}\]
\caption{Defining $\delta(\tuple{\aloc_1, \aloc_2})$ from
$\delta_1(\aloc_1)$ (rows) and $\delta_2(\aloc_2)$ (columns)}
\label{f:int.1NRA}
\end{figure}

\subsection{Counter Automata}
%\label{section-CA}

We define below two kinds of automata with counters,
namely without errors and with incrementing errors,
and then consider the complexity of deciding their nonemptiness,
over finite and over infinite words.

\subsubsection*{Automata}

A \emph{counter automaton} (CA) $\acaut$,
with $\emptyword$ transitions and zero testing,
is a tuple of the form
$\tuple{\aalphabet, \locs, \aloc_I, n, \delta, F}$, where:
\begin{itemize}
\item
$\aalphabet$ is a finite alphabet;
\item
$\locs$ is a finite set of locations;
\item
$\aloc_I$ is the initial location;
\item
$n \in \mathbb{N}$ is the number of counters;
\item
$\delta \subseteq
 \locs \times (\aalphabet \uplus \{\emptyword\}) \times L \times \locs$
is a transition relation over the instruction set
$L = \{\mathtt{inc, dec, ifz}\} \times \{1, \ldots, n\}$;
\item
$F \subseteq \locs$ is the set of accepting locations,
such that $\aloc' \notin F$ whenever
$\tuple{\aloc, \emptyword, l, \aloc'} \in \delta$.
\end{itemize}

A state of $\acaut$ is a pair $\tuple{\aloc, \acval}$
consisting of a location $\aloc \in \locs$ and a counter valuation
$\acval: \{1, \ldots, n\} \rightarrow \mathbb{N}$.

If $\acaut$ is \emph{Minsky} (i.e.\ without errors),
its transitions are of the form
$\tuple{\aloc, \acval} \stackrel{\aword, l}{\longrightarrow}
 \tuple{\aloc', \acval'}$,
which means that $\tuple{\aloc, \aword, l, \aloc'} \in \delta$
and $\acval'$ is obtained from $\acval$
by performing instruction $l$ in the standard manner,
where $l = \tuple{\mathtt{dec}, c}$ requires $\acval(c) > 0$
and $l = \tuple{\mathtt{ifz}, c}$ requires $\acval(c) = 0$.
A run of $\acaut$ is then a nonempty sequence of transitions
$\tuple{\aloc_0, \acval_0} \stackrel{\aword_0, l_0}{\longrightarrow}
 \tuple{\aloc_1, \acval_1} \stackrel{\aword_1, l_1}{\longrightarrow}
 \cdots$
where the initial state is given by
$\aloc_0 = \aloc_I$ and $\acval_0(c) = 0$ for each $c$.
We consider a finite run accepting iff it ends with an accepting location,
and an infinite run accepting iff accepting locations occur infinitely often.
We say that $\acaut$ accepts a word $\aword'$ over $\aalphabet$ iff
there exists a run as above which is accepting and such that
$\aword' = \aword_0 \aword_1 \ldots$.

The other case we consider is when $\acaut$ is \emph{incrementing},
i.e.\ its counters may erroneously increase at any time.
For counter valuations $\acval$ and $\acval_\dag$,
we write $\acval \leq \acval_\dag$ iff
$\acval(c) \leq \acval_\dag(c)$ for each $c$.
Runs and acceptance of incrementing $\acaut$
are defined in the same way as above,
but using transitions of the form
$\tuple{\aloc, \acval} \stackrel{\aword, l}{\longrightarrow}_\dag
 \tuple{\aloc', \acval'}$,
which means that there exist $\acval_\dag$ and $\acval'_\dag$ such that
$\acval \leq \acval_\dag$,
$\tuple{\aloc, \acval_\dag} \stackrel{\aword, l}{\longrightarrow}
 \tuple{\aloc', \acval'_\dag}$ and
$\acval'_\dag \leq \acval'$.
When it is clear that we are considering an incrementing CA,
we may write simply $\longrightarrow$ instead of $\longrightarrow_\dag$.

\begin{example}
Let $\acaut^{< \omega}$ be the $1$-counter automaton
with alphabet $\{\aletter, \aletterbis\}$
that is shown in Figure~\ref{f:CA.fin},
and $\acaut^\omega$ the $2$-counter automaton
with the same alphabet that is given by Figure~\ref{f:CA.inf},
where $\mathtt{ifnz}$ is used as syntactic sugar
for a decrement succeeded by an $\emptyword$ increment.
Since we shall consider acceptance of only finite words by $\acaut^{< \omega}$
and acceptance of only infinite words by $\acaut^\omega$,
their accepting locations are indicated in corresponding styles.

We have that $\acaut^{< \omega}$ (resp., $\acaut^\omega$) accepts
a finite (resp., infinite) word $\aword$ iff
every occurence of $\aletter$ is followed by
a separate occurence of $\aletterbis$,
which is iff there exists a data word $\adataword$
such that $\mathrm{str}(\adataword) = \aword$ and
which satisfies the LTL$^\downarrow_1(\nextt, \sometimes)$ sentence
$\aformula$ from Example~\ref{ex:LTL}.
That is the case regardless of whether
the automata are regarded as Minsky or incrementing.

Counter automata $\acaut^{< \omega}$ and $\acaut^\omega$ were obtained
from a register automaton $\aregaut$ as in Example~\ref{ex:RA},
using the proof of Theorem~\ref{th:register2counter} below.
\end{example}

\begin{narrowfig}{.62\textwidth}
\setlength{\unitlength}{4em}
\begin{picture}(6.5,2)(.25,1)
\gasset{Nw=.4,Nh=.4,Nmr=.2,Nadjust=n,loopCW=y}
\node[Nmarks=if,iangle=180,fangle=90](z)(1,2){}
\node(i)(3,2){}
\node(n)(5,2){}
\drawqbpedge(z,20,n,135){$\aletter, \mathtt{inc}, 1$}
\drawloop[loopdiam=.5,loopangle=270](z){$\aletterbis, \mathtt{ifz}, 1$}
\drawloop[loopdiam=.5,loopangle=0](n){$\aletter, \mathtt{inc}, 1$}
\drawloop[loopdiam=.5,loopangle=270](n){$\aletterbis, \mathtt{ifnz}, 1$}
\drawedge[ELside=r](n,i){$\emptyword, \mathtt{dec}, 1$}
\drawedge(i,z){$\aletterbis, \mathtt{ifz}, 1$}
\drawedge[curvedepth=-.3,ELside=r](i,n){$\aletterbis, \mathtt{ifnz}, 1$}
\end{picture}
\caption{A counter automaton over finite words}
\label{f:CA.fin}
\end{narrowfig}

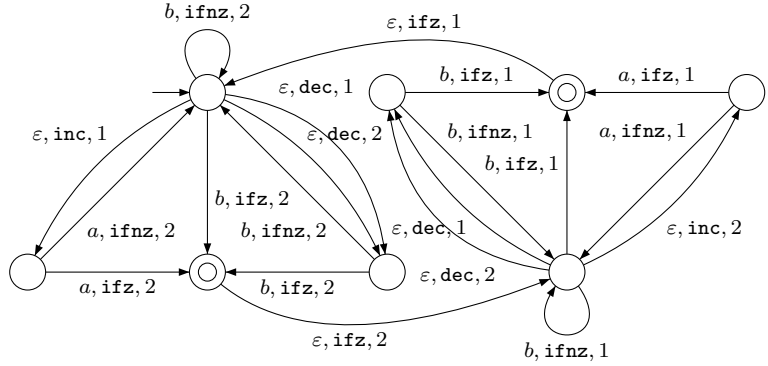
\begin{figure}
\setlength{\unitlength}{4em}
\begin{center}
\begin{picture}(8.5,4)(.75,1)
\gasset{Nw=.4,Nh=.4,Nmr=.2,Nadjust=n,loopCW=y}
\node[Nmarks=i,iangle=180](1)(3,4){}
\node(1i)(1,2){}
\node[Nmarks=r,rdist=.1](12)(3,2){}
\node(1d)(5,2){}
\node(2)(7,2){}
\node(2i)(9,4){}
\node[Nmarks=r,rdist=.1](21)(7,4){}
\node(2d)(5,4){}
\drawedge[curvedepth=-.3,ELside=r](1,1i){$\emptyword, \mathtt{inc}, 1$}
\drawedge[ELside=r,ELpos=40](1i,1){$\aletter, \mathtt{ifnz}, 2$}
\drawedge[ELside=r](1i,12){$\aletter, \mathtt{ifz}, 2$}
\drawedge[curvedepth=.6,ELpos=33](1,1d){$\emptyword, \mathtt{dec}, 1$}
\drawedge[curvedepth=.3,ELpos=50](1,1d){$\emptyword, \mathtt{dec}, 2$}
\drawedge[ELpos=40](1d,1){$\aletterbis, \mathtt{ifnz}, 2$}
\drawedge(1d,12){$\aletterbis, \mathtt{ifz}, 2$}
\drawloop[loopdiam=.5,loopangle=90](1){$\aletterbis, \mathtt{ifnz}, 2$}
\drawedge[ELpos=60](1,12){$\aletterbis, \mathtt{ifz}, 2$}
\drawqbpedge[ELside=r](12,135,2,22.5){$\emptyword, \mathtt{ifz}, 2$}
\drawedge[curvedepth=-.3,ELside=r](2,2i){$\emptyword, \mathtt{inc}, 2$}
\drawedge[ELside=r,ELpos=40](2i,2){$\aletter, \mathtt{ifnz}, 1$}
\drawedge[ELside=r](2i,21){$\aletter, \mathtt{ifz}, 1$}
\drawedge[curvedepth=.6,ELpos=33](2,2d){$\emptyword, \mathtt{dec}, 2$}
\drawedge[curvedepth=.3,ELpos=50](2,2d){$\emptyword, \mathtt{dec}, 1$}
\drawedge[ELpos=40](2d,2){$\aletterbis, \mathtt{ifnz}, 1$}
\drawedge(2d,21){$\aletterbis, \mathtt{ifz}, 1$}
\drawloop[loopdiam=.5,loopangle=270](2){$\aletterbis, \mathtt{ifnz}, 1$}
\drawedge[ELpos=60](2,21){$\aletterbis, \mathtt{ifz}, 1$}
\drawqbpedge[ELside=r](21,135,1,22.5){$\emptyword, \mathtt{ifz}, 1$}
\end{picture}
\end{center}
\caption{A counter automaton over infinite words}
\label{f:CA.inf}
\end{figure}

\subsubsection*{Complexity of Nonemptiness}

It turns out that CA with incrementing errors are easier to analyse
than CA without errors.

\begin{theorem}
\label{th:CA}
\begin{itemize}
\item[(a)]
For Minsky CA, nonemptiness is
$\Sigma^0_1$-complete over finite words,
and $\Sigma^1_1$-complete over infinite words.
\item[(b)]
For incrementing CA, nonemptiness is
decidable and not primitive recursive over finite words,
and $\Pi^0_1$-complete over infinite words.
\end{itemize}
\end{theorem}

\begin{proof}
For the finitary part of (a), we refer the reader to \cite{Minsky67},
and for the infinitary part, e.g.\ to \cite[Lemma~8]{Alur&Henzinger94}.
Note that the lower bounds hold already with
singleton alphabets, no $\emptyword$ transitions and $2$ counters.
Over finite words, those restrictions can be tightened by adding determinism,
so that for each location,
either there is one outgoing transition
and it is an increment,
or there are two outgoing transitions
and they are a decrement and a zero test of the same counter.

To obtain the finitary part of (b),
we observe that by reversing transition relations,
there are logarithmic-space reductions between
nonemptiness for incrementing CA over finite words and
reachability for classic lossy counter machines \cite{Mayr03}.
The latter problem is indeed decidable \cite[Theorem~6]{Mayr03}
and not primitive recursive \cite{Schnoebelen02}.

Incrementing CA can be seen as
insertion channel machines with emptiness testing (ICMETs)
\cite{Ouaknine&Worrell06a} whose message sets are singletons,
so $\Pi^0_1$-membership of nonemptiness for incrementing CA over infinite words
is a corollary of $\Pi^0_1$-membership of the recurrent-state problem for ICMETs
\cite{Ouaknine&Worrell06c}.
It can also be shown directly by considering the following procedure.
Given an incrementing CA $\acaut$, compute a tree of all states
which are reachable from the initial state.
Allowing only incrementing errors which are decrements
which do not alter the counter value makes the tree finitely branching.
Along every branch, stop as soon as
a state $\tuple{\aloc, \acval}$ is reached such that
either $\aloc$ is accepting or some previous state on the branch
is of the form $\tuple{\aloc, \acval'}$ with $\acval' \leq \acval$.
By Dickson's Lemma \cite{Dickson13}, each branch is finite,
so by K\"onig's Lemma, the computation terminates.
The above is then repeated from each leaf whose location is accepting.
It remains to observe that the procedure terminates iff
$\acaut$ does not have an accepting infinite run.

That nonemptiness for incrementing CA over infinite words is $\Pi^0_1$-hard
is obtained by verifying that the proof of $\Pi^0_1$-hardness of
the recurrent-state problem for ICMETs \cite[Theorem~2]{Ouaknine&Worrell06a}
can be adapted to the more restrictive setting of incrementing CA.
We reduce from emptiness over finite words for deterministic Minsky CA with
singleton alphabets, no $\emptyword$ transitions and $2$ counters.
Given such an automaton $\acaut$,
an incrementing CA $\widehat{\acaut}$
with $5$ counters $C_1$, $C_2$, $C'$, $D$ and $D'$,
which performs the pseudo-code in Figure~\ref{f:Pi01.hardness},
and whose unique accepting location corresponds to
the end of the \textbf{repeat} loop,
is computable in logarithmic space.
In the simulations of $\acaut$ by $\widehat{\acaut}$,
counter $D'$ prescribes how much ``space'' is allowed for
the two counters and the number of steps of $\acaut$.
As in the proof of \cite[Theorem~2]{Ouaknine&Worrell06a},
we have that $\widehat{\acaut}$ has an infinite accepting run iff
the unique infinite run of $\acaut$ does not contain an accepting location.
\end{proof}

\begin{narrowfig}{.67\textwidth}
\begin{tabbing}
\textbf{repeat} \\
\{ \= $D' := D$; \\
   \> \textbf{while} $D' > 0$ \\
   \> \{ \= simulate $\acaut$ using $C_1$ and $C_2$ as follows: \\
   \>    \> - if $\acaut$ accepts, $\widehat{\acaut}$ stops \\
   \>    \> - whenever $C_1$ or $C_2$ is decremented, increment $D'$ \\
   \>    \> - whenever $C_1$ or $C_2$ is incremented, decrement $D'$ \\
   \>    \> - after each step of $\acaut$, increment $C'$ and decrement $D'$ \\
   \>    \> - if $D' = 0$, exit the simulation; \\
   \>    \> $D' = D' + C_1 + C_2 + C' - 1$; $C_1, C_2, C' := 0$ \}; \\
   \> $D := D + 1$ \}
\end{tabbing}
\caption{Pseudo-code for $\widehat{\acaut}$}
\label{f:Pi01.hardness}
\end{narrowfig}

\subsection{Languages, Satisfiability and Nonemptiness}
%\label{section-lang-dec}

For a sentence $\aformula$ of
LTL$^\downarrow(\nextt, \nextt^{-1}, \until, \since)$ or
FO$(\sim, <, +1, \ldots, +m)$
with alphabet $\aalphabet$,
let $\mathrm{L}^{< \omega}(\aformula)$ (resp., $\mathrm{L}^\omega(\aformula)$)
denote the language of all finite (resp., infinite) data words over $\aalphabet$
which satisfy $\aformula$.
We say that $\aformula$ is satisfiable over finite or infinite data words iff
the corresponding language is nonempty.

Languages and nonemptiness of register automata and counter automata
are defined analogously.

\section{LTL with $1$ Register versus FO with $2$ Variables}
\label{s:LTL.FO}

It was proved in \cite{Etessami&Vardi&Wilke02} that
FO with $2$ variables and predicates $<$ and $+1$
is as expressive as unary LTL
(i.e., with operators $\nextt$, $\nextt^{-1}$,
$\sometimes$ and $\pastsometimes$),
but that in the worst case,
the latter is exponentially less succinct.
We now establish a similar equiexpressivness result for
FO$^2(\sim, <, +1, \ldots, +m)$.
First, we define the corresponding fragment of
LTL$^\downarrow(\nextt, \nextt^{-1}, \sometimes, \pastsometimes)$,
which has $1$ register and in which scopes of the freeze quantifier
are carefully restricted.

Suppose $m \in \mathbb{N}$.
Let $\mathcal{O}_m$ denote the following set of temporal operators:
\[\{\nextt, \nextt^{-1}, \ldots, \nextt^m, \nextt^{-m},
    \nextt^{m+1} \sometimes, \nextt^{-(m+1)} \pastsometimes\}\]
where $\nextt^k$ (resp., $\nextt^{-k}$) stands for
$k$ repetitions of $\nextt$ (resp., $\nextt^{-1}$).
We say that a formula of LTL$^\downarrow_1(\mathcal{O}_m)$ is \emph{simple} iff
each occurrence of a temporal operator is immediately preceded by $\downarrow_1$
(and there are no other occurences of $\downarrow_1$).

A sentence $\aformula$ of
LTL$^\downarrow(\nextt, \nextt^{-1}, \until, \since)$
is said to be equivalent to
a formula $\aformula'(x)$ of
FO$(\sim, <, +1, \ldots, +m)$ iff
they have the same alphabet $\aalphabet$ and,
for every data word $\adataword$ over $\aalphabet$
and $0 \leq i < \length{\adataword}$, we have
$\adataword, i \,\models_{\emptyset}\, \aformula
 \:\Leftrightarrow\:
 \adataword \models_{[x \mapsto i]} \aformula'(x)$.

\begin{example}
It is straightforward to rewrite
the sentence $\aformula$ from Example~\ref{ex:LTL}
so that it belongs to simple LTL$^\downarrow_1(\mathcal{O}_0)$.
Alternatively, that $\aformula$ is equivalent to
a sentence of simple LTL$^\downarrow_1(\mathcal{O}_0)$
is a consequence of the following theorem,
since it was observed in Example~\ref{ex:FO} that
$\aformula$ is equivalent to the formula $\aformula'(x_0)$ of FO$^2(\sim, <)$.
\end{example}

\begin{theorem}
\label{th:simple.FO2}
\begin{itemize}
\item[(a)]
For each sentence of simple LTL$^\downarrow_1(\mathcal{O}_m)$,
an equivalent formula of FO$^2(\sim, <, +1, \ldots, +m)$
is computable in logarithmic space.
\item[(b)]
For each formula $\aformula(x_j)$ of FO$^2(\sim, <, +1, \ldots, +m)$,
an equivalent sentence of simple LTL$^\downarrow_1(\mathcal{O}_m)$
is computable in polynomial space.
\end{itemize}
\end{theorem}

\begin{proof}
The following notations will be convenient.
Let $\atempop^0 = \downarrow_1$,
$\atempop^k = {\downarrow_1} \nextt^k$ for
$k \in \{-m, \ldots, -1, 1, \ldots, m\}$,
$\atempop^{m+1} = {\downarrow_1} \nextt^{m+1} \sometimes$, and
$\atempop^{-(m+1)} = {\downarrow_1} \nextt^{-(m+1)} \pastsometimes$.
For $j \in \{0, 1\}$, let
\[\begin{array}{rcl}
\aformulater^j_0 & \egdef &
x_{1-j} = x_j \\
\aformulater^j_k & \egdef &
x_{1-j} = x_j + k \ (1 \leq k \leq m) \\
\aformulater^j_{-k} & \egdef &
x_j = x_{1 - j} + k \ (1 \leq k \leq m) \\
\aformulater^j_{m+1} & \egdef &
x_{j} < x_{1-j} \wedge \bigwedge_{1 \leq k \leq m} \neg x_{1-j} = x_{j} + k \\
\aformulater^j_{-(m+1)} & \egdef &
x_{1-j} < x_{j} \wedge \bigwedge_{1 \leq k \leq m} \neg x_{j} = x_{1-j} + k
\end{array}\]
(The equality predicate can be expressed using $<$.)

We have (a) by the following translations $T_j$ which map
simple LTL$^\downarrow_1(\mathcal{O}_m)$ formulae to
FO$^2(\sim, <, +1, \ldots, +m)$ formulae.
Each sentence $\aformula$ will be equivalent to $T_j(\aformula)$
which will contain at most $x_j$ free.
The maps $T_j$ are defined by structural recursion,
by encoding the semantics of simple formulae into first-order logic,
and by recycling variables (to use only two variables).
The Boolean clauses are omitted.
\[
T_j(\aletter) \egdef
P_{\aletter}(x_j)
\hspace{2em}
T_j(\uparrow_1) \egdef
x_{1 - j} \sim x_j
\hspace{2em}
T_j(\atempop^k \aformulabis) \egdef
\eexists{x_{1 - j}}
{\aformulater^j_k \wedge T_{1 - j}(\aformulabis)}
\]

For (b), we proceed by adapting the proof of
\cite[Theorem~1]{Etessami&Vardi&Wilke02}.
We define recursively translations $T'_j$
from FO$^2(\sim, <, +1, \ldots, +m)$ formulae $\aformula(x_j)$
to equivalent simple LTL$^\downarrow_1(\mathcal{O}_m)$ sentences.
The cases of Boolean operators and one-variable atomic formulae
are straightforward.
The remaining case is when $\aformula(x_j)$ is of the form
\[
\exists x_{1-j} \,
\beta(\alpha_1(x_0, x_1), \ldots, \alpha_L(x_0, x_1),
      \xi_1(x_j), \ldots, \xi_N(x_j),
      \zeta_1(x_{1-j}), \ldots, \zeta_M(x_{1-j}))
\]
where $\beta$ is a Boolean formula,
and each $\alpha_i(x_0, x_1)$ is a $\sim$, $<$ or $+k$ atomic formula.
Now, for each $-(m+1) \leq k \leq m+1$ and $b \in \{\top, \bot\}$,
let $\alpha_i^{k, b}$ denote the truth value of $\alpha_i(x_0, x_1)$
under assumptions $\aformulater^j_k$ and
$x_j \sim x_{1-j} \,\Leftrightarrow\, b$.
Also, for each $\aset \subseteq \{1, \ldots, N\}$,
let $\xi_i^\aset = \top$ if $i \in \aset$,
and $\xi_i^\aset = \bot$ otherwise.
$T'_j(\aformula(x_j))$ is then computed as
\[\begin{array}{c}
\bigvee_{-(m+1) \leq k \leq m+1}
\bigvee_{b \in \{\top, \bot\}}
\bigvee_{\aset \subseteq \{1, \ldots, N\}}
\big(\bigwedge_{i \in \{1, \ldots, N\}}
     T'_j(\xi_i(x_j)) \Leftrightarrow \xi_i^\aset\big) \wedge \\
\atempop^k (({\uparrow_1} \Leftrightarrow b) \wedge
            \beta(\alpha_1^{k, b}, \ldots, \alpha_L^{k, b},
                  \xi_1^\aset, \ldots, \xi_N^\aset,
                  T'_{1-j}(\zeta_1(x_{1-j})), \ldots,
                  T'_{1-j}(\zeta_M(x_{1-j})))
\end{array}\]

The size of the equivalent simple LTL$^\downarrow_1(\mathcal{O}_m)$ formula
is exponential in $\length{\aformula}$,
because the length of the stack of recursive calls is linear
and the generalised conjunctions and disjunctions
have at most exponentially many arguments.
For the same reasons, polynomial space is sufficient for the computation.
\end{proof}

It was shown in \cite{Bojanczyketal06a} that,
over finite data words,
satisfiability for FO$^2(\sim, <, +1)$
is reducible in doubly exponential time to
reachability for Petri nets,
and that there is a polynomial-time reduction in the reverse direction.
Reachability for Petri nets is known to be
decidable (cf.\ e.g.\ \cite{Kosaraju82}) and
\textsc{ExpSpace}-hard \cite{Lipton76}.
Two extensions of the decidability of satisfiability for FO$^2(\sim, <, +1)$
over finite data words were also obtained in \cite{Bojanczyketal06a}:
for FO$^2(\sim, <, +1, \ldots, +m)$, and over infinite data words.
By the following corollary of Theorem~\ref{th:simple.FO2},
those results have immediate consequences for
complexity of satisfiability problems for
simple fragments of LTL$^\downarrow_1(\mathcal{O}_m)$.

\begin{corollary}
\label{cor:simple.FO2}
Over finite and over infinite data words,
satisfiability for the simple fragment of LTL$^\downarrow_1(\mathcal{O}_m)$
is reducible in logarithmic space to
satisfiability for FO$^2(\sim, <, +1, \ldots, +m)$,
and there is a polynomial-space reduction in the reverse direction.
\end{corollary}

\section{Upper Complexity Bounds}
\label{section-upper-bounds}

A number of upper bounds on
complexity of satisfiability for
fragments of LTL with the freeze quantifier and
complexity of nonemptiness for
classes of register automata
will be obtained below.
The former will be corollaries of the latter,
by the following result which shows that
logical sentences are easily translatable to equivalent automata.
Note that, in contrast to classical automata,
alternating register automata are more expressive than
nondeterministic and universal,
and two-way register automata are more expressive than one-way
\cite{Kaminski&Francez94,Neven&Schwentick&Vianu04}.
As a specific example, the ``nonces property'' that
no two word positions are in the same class is expressible
in LTL$^\downarrow_1(\nextt, \sometimes)$ (cf.\ Example~\ref{ex:LTL})
and by an automaton in 1ARA$_1$,
but not by any automaton in 1NRA.

\begin{theorem}
\label{th:logic2register}
For each sentence $\aformula$ of
LTL$^\downarrow_n(\nextt, \nextt^{-1}, \until, \since)$,
an automaton $\aregaut_\aformula$ in 2ARA$_n$
with the same alphabet and such that
$\mathrm{L}^{< \omega}(\aformula) =
 \mathrm{L}^{< \omega}(\aregaut_\aformula)$ and
$\mathrm{L}^\omega(\aformula) =
 \mathrm{L}^\omega(\aregaut_\aformula)$
is computable in logarithmic space.
If $\aformula$ is in LTL$^\downarrow_n(\nextt, \until)$,
$\aregaut_\aformula$ is in 1ARA$_n$.
\end{theorem}

\begin{proof}
The translation is a simple extension of
the classical one from LTL to alternating automata
(cf.\ e.g.\ \cite{Vardi96}).

We can assume that $\aformula$ is in negation normal form,
where we write
$\overline{\aletter}$,
$\bot$,
$\vee$,
$\overline{\atempop}$ and
$\nuparrow_r$
for the duals of
$\aletter$,
$\top$,
$\wedge$,
$\atempop \in \{\nextt, \nextt^{-1}, \until, \since\}$ and
$\uparrow_r$.

Let $\mathrm{cl}(\aformula)$ be the set of all subformulae of $\aformula$,
together with $\top$, $\bot$, and all subformulae of
$\aformulabis \wedge
 \nextt (\aformulabis \until \aformulater)$,
$\aformulabis \wedge
 \nextt^{-1} (\aformulabis \since \aformulater)$,
$\aformulabis \vee
 \overline{\nextt} (\aformulabis \overline{\until} \aformulater)$ or
$\aformulabis \vee
 \overline{\nextt^{-1}} (\aformulabis \overline{\since} \aformulater)$
for each subformula of $\aformula$ of the form
$\aformulabis \until \aformulater$,
$\aformulabis \since \aformulater$,
$\aformulabis \overline{\until} \aformulater$ or
$\aformulabis \overline{\since} \aformulater$
(respectively).

To define
$\aregaut_\aformula =
 \tuple{\aalphabet, \locs, \aloc_I, n, \delta, \rank, \height}$,
let
$\locs =
 \{\aloc_\aformulabis \,:\,
   \aformulabis \in \mathrm{cl}(\aformula)\}$
and
$\aloc_I = \aloc_\aformula$.

The transition function is given below,
where we omit dual cases:
\[\begin{array}{rcl@{\hspace{1.4em}}rcl@{\hspace{1.4em}}rcl}
\delta(\aloc_\aletter) & = &
\ifte{\aletter}{\aloc_\top}{\aloc_\bot}
&
\delta(\aloc_\top) & = &
\top
&
\delta(\aloc_{\aformulabis \wedge \aformulater}) & = &
\aloc_\aformulabis \wedge \aloc_\aformulater
\\
\delta(\aloc_{\uparrow_r}) & = &
\ifte{{\uparrow_r}}{\aloc_\top}{\aloc_\bot}
&
\delta(\aloc_{\nextt \aformulabis}) & = &
\nextt \aloc_\aformulabis
&
\delta(\aloc_{\aformulabis \until \aformulater}) & = &
\aloc_\aformulater \vee
\aloc_{\aformulabis \wedge \nextt (\aformulabis \until \aformulater)}
\\
\delta(\aloc_{{\downarrow_r} \aformulabis}) & = &
{\downarrow_r} \aloc_\aformulabis
&
\delta(\aloc_{\nextt^{-1} \aformulabis}) & = &
\nextt^{-1} \aloc_\aformulabis
&
\delta(\aloc_{\aformulabis \since \aformulater}) & = &
\aloc_\aformulater \vee
\aloc_{\aformulabis \wedge \nextt^{-1} (\aformulabis \since \aformulater)}
\end{array}\]

The ranks are defined so that
every $\aloc_{\aformulater \until \aformulater'}$ has odd rank and
every $\aloc_{\aformulater \overline{\until} \aformulater'}$ has even rank.
For example, $\rank(\aloc_\aformulabis) = 2 \length{\aformulabis}$
unless $\aformulabis$ is of the form $\aformulater \until \aformulater'$,
in which case $\rank(\aloc_\aformulabis) = 2 \length{\aformulabis} + 1$.

The heights may be defined as
$\height(\aloc_\aformulabis) = \length{\aformulabis}$
unless $\aformulabis$ is of the form
$\aformulater \until \aformulater'$,
$\aformulater \since \aformulater'$,
$\aformulater \overline{\until} \aformulater'$ or
$\aformulater \overline{\since} \aformulater'$,
in which case
$\height(\aloc_{\aformulater \until \aformulater'}) =
 \length{\aformulater' \vee (\aformulater \wedge
           \nextt (\aformulater \until \aformulater'))}$
etc.

It remains to show the equalities between the languages of
$\aformula$ and $\aregaut_\aformula$,
so suppose $\adataword$ is a data word over $\aalphabet$.
By a straightforward induction on $\aformulabis \in \mathrm{cl}(\aformula)$,
it holds that, for every $0 \leq i < \length{\adataword}$
and $n$-register valuation $\aregval$ for $\adataword$,
we have $\adataword, i \,\models_\aregval\, \aformulabis$
iff player $1$ has a winning strategy
from state $\tuple{i, \aloc_\aformulabis, \aregval}$
in game $\agame_{\aregaut_\aformula, \adataword}$.
In particular, $\adataword, 0 \,\models_\emptyset\, \aformula$
iff $\aregaut_\aformula$ accepts $\adataword$.
\end{proof}

The following basic upper bounds should be compared with
the lower bounds in Theorem~\ref{theorem-the-last}.

\begin{theorem}
\label{th:basic-upper-bounds}
Over finite data words,
satisfiability for LTL$^\downarrow(\nextt, \nextt^{-1}, \until, \since)$
and nonemptiness for 2ARA are in $\Sigma^0_1$.
Over infinite data words,
satisfiability for LTL$^\downarrow(\nextt, \nextt^{-1}, \until, \since)$
and nonemptiness for 2ARA are in $\Sigma^1_1$,
and nonemptiness for 2NRA is in $\Sigma^0_2$.
\end{theorem}

\begin{proof}
By Theorem~\ref{th:logic2register}, it suffices to consider
the register automata nonemptiness problems.

That nonemptiness for 2ARA
is in $\Sigma^0_1$ over finite data words
and in $\Sigma^1_1$ over infinite data words
are straightforward consequences of Theorem~\ref{th:WG}.

Suppose $\aregaut$ is in 2NRA.
Because of nondeterminism, $\aregaut$ accepts a data word iff
there exists a complete play from the initial state in the acceptance game
which is winning for player $1$.
By K\"onig's Lemma, we have that $\aregaut$ accepts an infinite data word iff
there exists $j \in \mathbb{N}$ such that for each $k \in \mathbb{N}$:
\begin{describe}{(*)}
\item[(*)]
there exist a data word $\adataword$ of length $k + 1$
and a play $\aplay = \apos_0 \apos_1 \ldots$ of length at most $k$
from the initial state in $\agame_{\aregaut, \adataword}$ such that
either $\aplay$ is winning for player $1$,
or $\pi$ is of length $k$ and
$\rank(\apos_{j'})$ is even for all $j' \geq j$.
\end{describe}
The $\Sigma^0_2$-membership follows by observing that (*) is decidable.
\end{proof}

It was shown in \cite{Sakamoto&Ikeda00} that
nonemptiness for one-way nondeterministic register automata
over finite data words is in \textsc{NP},
and that the problem is \textsc{NP}-hard already for deterministic automata.
However, due to the technical differences noted in Remark~\ref{rem:RA},
the complexity for the notion of register automata in this paper
turns out to be \textsc{PSpace}-complete, over infinite data words as well.
The proof below of the \textsc{PSpace}-memberships
will also prepare us for establishing Theorem~\ref{th:register2counter}.
The hardness results are in Theorem~\ref{th:pspace-hard}.

We remark that, for infinite data words, it is straightforward to extend
Theorems \ref{th:pspace} and \ref{th:register2counter} to register automata
with B\"uchi acceptance, without affecting the complexity bounds.

\begin{theorem}
\label{th:pspace}
The following hold over finite and over infinite data words:
\begin{itemize}
\item
nonemptiness for 1NRA is in \textsc{PSpace};
\item
for every fixed $n \in \mathbb{N}$,
nonemptiness for 1NRA$_n$ is in \textsc{NLogSpace}.
\end{itemize}
\end{theorem}

\begin{proof}
Suppose
$\aregaut = \tuple{\aalphabet, \locs, \aloc_I, n, \delta, \rank, \height}$
is in 1NRA.

Let $H_\aregaut$ be the set of all ``abstract states'' of the form
$\tuple{\aletter, \mathit{ee}, R, \aloc, E}$ where
$\aletter \in \aalphabet$,
$\mathit{ee} \in \{\top, \bot\}$,
$R$ is either $\emptyset$ or a class of $E$,
$\aloc \in \locs$ and
$E$ is an equivalence relation on a subset of $\{1, \ldots, n\}$.
For a data word $\adataword$ (over $\aalphabet$),
let $\alpha_{\aregaut, \adataword}$ be the following mapping from
states of $\aregaut$ (for $\adataword$) to elements of $H_\aregaut$:
\[\begin{array}{r}
\alpha_{\aregaut, \adataword}(\tuple{i, \aloc, \aregval}) =
\tuple{\adataword(i), i + 1 = \length{\adataword},
       \{r \,:\, \aregval(r) = [i]_\sim\}, \aloc, \\
       \{\tuple{r, r'} \,:\, r, r' \in \mathrm{dom}(\aregval) \mathrm{\ and\ }
                             \aregval(r) = \aregval(r')\}}
\end{array}\]
We define a relation $\rightarrow$ on $H_\aregaut$ by $h \rightarrow h'$ iff
there exist a data word $\adataword$ and
states $\apos$ and $\apos'$ of $\aregaut$ such that
$\alpha_{\aregaut, \adataword}(\apos) = h$,
$\alpha_{\aregaut, \adataword}(\apos') = h'$ and
$\apos \rightarrow \apos'$.
We say that $h \in H_\aregaut$ is
\emph{initial} (resp., \emph{winning}) iff
for some (equivalently, for every)
data word $\adataword$ and state $\apos$ of $\aregaut$
such that $\alpha_{\aregaut, \adataword}(\apos) = h$,
we have that $\apos$ is
initial (resp., has no successors and belongs to player $2$).

Because of nondeterminism, $\aregaut$ accepts a data word $\adataword$ iff
there exists a complete play from the initial state in
$\agame_{\aregaut, \adataword}$ which is winning for player $1$.
Since $\aregaut$ is one-way, for every sequence
$h_0 \rightarrow h_1 \rightarrow \cdots$ in $H_\aregaut$ with $h_0$ initial,
there exist a data word $\adataword$ and a play
$\apos_0 \apos_1 \ldots$ from the initial state in
$\agame_{\aregaut, \adataword}$ such that
$\alpha_{\aregaut, \adataword}(\apos_j) = h_j$ for all $j$.
Consequently:
\begin{itemize}

\item
$\mathrm{L}^{< \omega}(\aregaut)$ is nonempty iff there exists a sequence
$h_0 \rightarrow h_1 \rightarrow \cdots h_k$ in $H_\aregaut$
such that $h_0$ is initial and $h_k$ is winning;

\item
$\mathrm{L}^\omega(\aregaut)$ is nonempty iff:
\begin{itemize}
\item
either there exists a sequence
$h_0 \rightarrow h_1 \rightarrow \cdots h_k$ in $H_\aregaut$
such that $h_0$ is initial, $h_k$ is winning,
and the second component of $h_k$ is $\bot$,
\item
or there exists a sequence
$h_0 \rightarrow h_1 \rightarrow \cdots
 h_k \rightarrow h_{k + 1} \rightarrow \cdots h_{k'}$ in $H_\aregaut$
such that $h_0$ is initial, $h_k = h_{k'}$,
and the rank of the location in $h_k$ and $h_{k'}$ is even.
\end{itemize}
\end{itemize}

It remains to observe that,
for storing an abstract state and
for checking the successor relation
and the initial and winning properties on abstract states,
space which is logarithmic in $\length{\aalphabet}$ and $\length{\locs}$
and polynomial in $n$ suffices.
\end{proof}

In terms of the definitions in this paper,
it was established in \cite[Appendix~A]{Kaminski&Francez94}
that containment of the language of an automaton in 1NRA
in the language of an automaton in 1NRA$_1$
is decidable over finite data words.
In particular, nonemptiness for 1URA$_1$ (see Theorem~\ref{th:closure})
over finite data words is decidable.
We now prove the main result of this section,
which shows that decidability in fact holds for 1ARA$_1$,
and therefore also for LTL$^\downarrow_1(\nextt, \until)$ satisfiability.
The two problems over infinite data words are shown to be co-r.e.
The proof is by reductions to nonemptiness of incrementing counter automata,
which will provide the first half of a correspondence between
languages of incrementing CA and sentences of future-time fragments
of LTL with $1$ register (see Corollary~\ref{cor:languages}).

Using the developments in the proof of Theorem~\ref{th:pspace},
it is straightforward to extend the argument below
to obtain the same upper bounds for
containment of the language of an automaton in 1NRA
in the language of an automaton in 1ARA$_1$,
or in the language of a sentence in LTL$^\downarrow_1(\nextt, \until)$.
Over infinite data words, extending further to
one-way nondeterministic register automata with B\"uchi acceptance
requires no extra work.

Theorem~\ref{theorem-the-last} will show that
decidability and $\Pi^0_1$-membership break down as soon as
any of $1$ more register, past-time operators or backward moves are added.

\begin{theorem}
\label{th:register2counter}
Satisfiability for LTL$^\downarrow_1(\nextt, \until)$ and
nonemptiness for 1ARA$_1$ are
decidable over finite data words,
and in $\Pi^0_1$ over infinite data words.
\end{theorem}

\begin{proof}
By Theorems \ref{th:logic2register} and \ref{th:CA} (b),
it suffices to show that, given $\aregaut$ in 1ARA$_1$,
incrementing CA $\acaut_\aregaut^{< \omega}$ and $\acaut_\aregaut^{< \omega}$
are computable such that
\[\mathrm{L}^{< \omega}(\acaut_\aregaut^{< \omega}) =
  \{\mathrm{str}(\adataword) \,:\,
    \adataword \in \mathrm{L}^{< \omega}(\aregaut)\}
  \hspace{2em}
  \mathrm{L}^\omega(\acaut_\aregaut^\omega) =
  \{\mathrm{str}(\adataword) \,:\,
    \adataword \in \mathrm{L}^\omega(\aregaut)\}\]
In both cases, the proof will consist of the following steps:
\begin{itemize}
\item
replace the two-player acceptance games for $\aregaut$ by one-player games
whose positions are built from sets of states of $\aregaut$
($\aregaut$ cannot in general be translated to an automaton in 1NRA:
see the remarks which precede Theorem~\ref{th:logic2register}),
and whose successors are ``big step'' in the sense that they correspond to
following strategies for the automaton until
first moves to the next word position;
\item
combine the one-player acceptance games
with searching for a data word to be accepted,
resulting in a one-player nonemptiness game for $\aregaut$;
\item
show how to construct a CA which guesses and checks
a winning play in the nonemptiness game;
\item
show that allowing incrementing errors in computations of the CA
does not increase its language (such errors in an accepting computation
will amount to introducing superfluous states of $\aregaut$ from which
winning strategies for the automaton are then found).
\end{itemize}

First, we consider computing $\acaut_\aregaut^{< \omega}$.
Let $\aregaut = \tuple{\aalphabet, \locs, \aloc_I, 1, \delta, \rank, \height}$.

To define a big-step successor relation between sets of states,
for a state $\apos$ and a set of states $\poss'$
of $\aregaut$ for a data word $\adataword$ over $\aalphabet$,
let us write $\apos \Rightarrow \poss'$ iff
there exists a strategy $\astr$ for player $1$ from $\apos$
in game $\agame_{\aregaut, \adataword}$ such that:
\begin{itemize}
\item
each complete play $\aplay \in \astr$
which contains no move to another word position
is winning for player $1$;
\item
$\poss'$ is the set of all targets
of first moves to another word position
in plays of $\astr$.
\end{itemize}
For sets of states $\poss \neq \emptyset$ and $\poss'$,
we write $\poss \Rightarrow \poss'$ iff
there exists a map $\apos \mapsto \poss'_\apos$ on $\poss$ such that
$\apos \Rightarrow \poss'_\apos$ for each $\apos \in \poss$ and
$\poss' = \bigcup_{\apos \in \poss} \poss'_\apos$.
Since $\aregaut$ is one-way,
if the first component of every state in $\poss$ is $i$
and $\poss \Rightarrow \poss'$,
then the first component of every state in $\poss'$ is $i + 1$.
We call a set of states \emph{unipositional} iff
its members have the same first components.

By positional determinacy of weak games (see Theorem~\ref{th:WG}),
the decreasing heights discipline of register automata,
and K\"onig's Lemma,
we have:
\begin{describe}{(I)}
\item[(I)]
for every finite data word $\adataword$ over $\aalphabet$,
$\aregaut$ accepts $\adataword$ iff
there exists a sequence $\poss_1$, \ldots, $\poss_{k - 1}$
of sets of states of $\aregaut$ for $\adataword$ such that
$\{\tuple{0, \aloc_I, \emptyset}\} \Rightarrow
 \poss_1 \Rightarrow \cdots \poss_{k - 1} \Rightarrow
 \emptyset$.
\end{describe}

Now, let $H_\aregaut$ consist of $\emptyset$
and all ``abstract sets'' of the form
$\tuple{\aletter, \mathit{ee}, \locs_=, \locs_\emptyset, \sharp}$
where
$\aletter \in \aalphabet$,
$\mathit{ee} \in \{\top, \bot\}$,
$\locs_=, \locs_\emptyset \subseteq \locs$,
$\sharp: \mathcal{P}(\locs) \setminus \{\emptyset\} \,\rightarrow\, \mathbb{N}$,
and either $\locs_= \neq \emptyset$ or $\locs_\emptyset \neq \emptyset$
or $\sharp(\locs_\dag) > 0$ for some $\locs_\dag$.
We define a mapping $\alpha_{\aregaut, \adataword}$
from unipositional sets of states of $\aregaut$ for $\adataword$
to elements of $H_\aregaut$ as follows:
$\alpha_{\aregaut, \adataword}(\emptyset) = \emptyset$,
and for nonempty $\poss$ whose members' first component is $i$,
\[\begin{array}{r}
\alpha_{\aregaut, \adataword}(\poss) =
\tuple{\adataword(i), i + 1 = \length{\adataword},
       \{\aloc \,:\, \tuple{i, \aloc, [1 \mapsto [i]_\sim]} \in \poss\},
       \{\aloc \,:\, \tuple{i, \aloc, \emptyset} \in \poss\}, \\
       \locs_\dag \mapsto \length{\{D \neq [i]_\sim \::\:
         \{\aloc \,:\, \tuple{i, \aloc, [1 \mapsto D]} \in \poss\}
         = \locs_\dag\}}}
\end{array}\]
In particular, the last component of
the abstract set $\alpha_{\aregaut, \adataword}(\poss)$
maps each nonempty $\locs_\dag \subseteq \locs$
to the number of distinct data $D$ which are not the class of $i$ and such that
the set of all $\aloc$ with $\tuple{i, \aloc, [1 \mapsto D]} \in \poss$
equals $\locs_\dag$.

As the first step to defining
a big-step successor relation between abstract sets,
for
$\aletter \in \aalphabet$,
$\mathit{ee}, \mathit{uu} \in \{\top, \bot\}$ and
$\aloc \in \locs$,
let $\Succ{\aletter, \mathit{ee}, \mathit{uu}, \aloc}$
be the set of pairs of sets of locations
that is defined in Figure~\ref{f:def.abs.big.succ}
by recursion over the height of $\aloc$.
(Observe that $\locs'_{\neq} = \emptyset$ whenever
$\tuple{\locs'_{\neq}, \locs'_=} \in
  \Succ{\aletter, \mathit{ee}, \top, \aloc}$.)
Those sets satisfy:
\begin{describe}{(II)}
\item[(II)]
for every data word $\adataword$ over $\aalphabet$,
state $\tuple{i, \aloc, \aregval}$ and set of states $\poss'$
of $\aregaut$ for $\adataword$,
we have $\tuple{i, \aloc, \aregval} \Rightarrow \poss'$ iff
there exists
\[\tuple{\locs'_{\neq}, \locs'_=} \in
  \Succ{\adataword(i), i + 1 = \length{\adataword},
        \aregval = [1 \mapsto [i]_\sim], \aloc}\]
such that
$\poss' =
 \{\tuple{i + 1, \aloc', \aregval} \,:\, \aloc' \in \locs'_{\neq}\} \,\cup\,
 \{\tuple{i + 1, \aloc', [1 \mapsto [i]_\sim]} \,:\, \aloc' \in \locs'_=\}$.
\end{describe}
The following notations will be useful: given a map
$\amap: \aset \,\rightarrow\,
        \mathcal{P}(\asetbis_1) \times \mathcal{P}(\asetbis_2)$,
let
\[\begin{array}{rcl@{\hspace{2em}}rcl}
\bigcup_1 \amap & = &
\bigcup \{\asetter_1 \,:\, \tuple{\asetter_1, \asetter_2} \in \amap(\aset)\} &
\bigcup_2 \amap & = &
\bigcup \{\asetter_2 \,:\, \tuple{\asetter_1, \asetter_2} \in \amap(\aset)\}
\end{array}\]
For $h, h' \in H_\aregaut$, we write $h \Rightarrow h'$ iff
$h$ is of the form
$\tuple{\aletter, \mathit{ee},
        \locs_=, \locs_\emptyset, \sharp}$
and there exist maps
$\aloc \in \locs_= \,\mapsto\, \amap_=(\aloc) \in
   \Succ{\aletter, \mathit{ee}, \top, \aloc}$,
$\aloc \in \locs_\emptyset \,\mapsto\, \amap_\emptyset(\aloc) \in
   \Succ{\aletter, \mathit{ee}, \bot, \aloc}$ and
$\aloc \in \locs_\dag \,\mapsto\, \amap_{\locs_\dag, j}(\aloc) \in
   \Succ{\aletter, \mathit{ee}, \bot, \aloc}$
for each nonempty $\locs_\dag \subseteq \locs$
and $j \in \{1, \ldots, \sharp(\locs_\dag)\}$
such that:
\begin{itemize}
\item
either $h' = \emptyset$, $\bigcup_1 \amap_\emptyset = \emptyset$,
and $\sharp'(\locs'_\dag) = 0$ for all $\locs'_\dag$,
\item
or $h'$ is of the form
$\tuple{\aletter', \mathit{ee}',
        \emptyset, \bigcup_1 \amap_\emptyset, \sharp'}$,
\item
or $h'$ is of the form
$\tuple{\aletter', \mathit{ee}',
        \locs'_=, \bigcup_1 \amap_\emptyset,
        \sharp'[\locs'_= \mapsto \sharp'(\locs'_=) - 1]}$,
\end{itemize}
where, for each nonempty $\locs'_\dag \subseteq \locs$,
$\sharp'(\locs'_\dag)$ is defined as
\[\length{\{\tuple{\locs_\dag, j} \,:\,
            \textstyle{\bigcup}_1 \amap_{\locs_\dag, j} = \locs'_\dag\}}
+ \left\{\begin{array}{ll}
  1, & \mathrm{if\ }
  \bigcup_2 \amap_= \,\cup\,
  \bigcup_2 \amap_\emptyset \,\cup\,
  \bigcup_{\locs_\dag, j} \bigcup_2 \amap_{\locs_\dag, j}
  = \locs'_\dag \\
  0, & \mathrm{otherwise}
  \end{array}\right.\]
From (II), it follows that:
\begin{describe}{(III)}
\item[(III)]
for every data word $\adataword$ over $\aalphabet$,
and unipositional sets $\poss$ and $\poss'$
of states of $\aregaut$ for $\adataword$,
we have $\poss \Rightarrow \poss'$ iff
$\alpha_{\aregaut, \adataword}(\poss) \Rightarrow
 \alpha_{\aregaut, \adataword}(\poss')$.
\end{describe}

\begin{figure}
\[\begin{array}{r|c|}
\delta(\aloc) &
\Succ{\aletter, \mathit{ee}, \mathit{uu}, \aloc}
\\ \hline
\ifte{\beta}{\aloc'}{\aloc''} &
\Succ{\aletter, \mathit{ee}, \mathit{uu}, \aloc'},
\mathrm{\ if\ }
\aletter, \mathit{ee}, \mathit{uu} \models \beta
\hspace{2em}
\Succ{\aletter, \mathit{ee}, \mathit{uu}, \aloc''},
\mathrm{\ if\ }
\aletter, \mathit{ee}, \mathit{uu} \not\models \beta
\\ \hline
{\downarrow_1} \aloc' &
\Succ{\aletter, \mathit{ee}, \top, \aloc'}
\\ \hline
\aloc' \wedge \aloc'' &
\{\tuple{\locs'_{\neq} \cup \locs''_{\neq}, \locs'_= \cup \locs''_=} \,:\,
  \tuple{\locs'_{\neq}, \locs'_=} \in
  \Succ{\aletter, \mathit{ee}, \mathit{uu}, \aloc'},
  \tuple{\locs''_{\neq}, \locs''_=} \in
  \Succ{\aletter, \mathit{ee}, \mathit{uu}, \aloc''}\}
\\ \hline
\aloc' \vee \aloc'' &
\Succ{\aletter, \mathit{ee}, \mathit{uu}, \aloc'} \,\cup\,
\Succ{\aletter, \mathit{ee}, \mathit{uu}, \aloc''}
\\ \hline
\top &
\{\tuple{\emptyset, \emptyset}\}
\\ \hline
\bot &
\emptyset
\\ \hline
\nextt \aloc' &
\begin{array}{c}
\{\tuple{\{\aloc'\}, \emptyset}\},
\mathrm{\ if\ }
\mathtt{ee} = \bot \mathrm{\ and\ } \mathit{uu} = \bot
\hspace{2em}
\{\tuple{\emptyset, \{\aloc'\}}\},
\mathrm{\ if\ }
\mathtt{ee} = \bot \mathrm{\ and\ } \mathit{uu} = \top
\\
\emptyset, \mathrm{\ if\ } \mathtt{ee} = \top
\end{array}
\\ \hline
\overline{\nextt} \aloc' &
\begin{array}{c}
\{\tuple{\{\aloc'\}, \emptyset}\},
\mathrm{\ if\ }
\mathtt{ee} = \bot \mathrm{\ and\ } \mathit{uu} = \bot
\hspace{2em}
\{\tuple{\emptyset, \{\aloc'\}}\},
\mathrm{\ if\ }
\mathtt{ee} = \bot \mathrm{\ and\ } \mathit{uu} = \top
\\
\{\tuple{\emptyset, \emptyset}\}, \mathrm{\ if\ } \mathtt{ee} = \top
\end{array}
\\ \hline
\end{array}\]
\[\begin{array}{rcl@{\hspace{2em}}rcl@{\hspace{2em}}rcl}
\aletter, \mathit{ee}, \mathit{uu} \models \aletter'
& \equivdef &
\aletter = \aletter'
&
\aletter, \mathit{ee}, \mathit{uu} \models \mathtt{end}
& \equivdef &
\mathit{ee} = \top
&
\aletter, \mathit{ee}, \mathit{uu} \models {\uparrow_1}
& \equivdef &
\mathit{uu} = \top
\end{array}\]
\caption{Defining abstract big-step successors}
\label{f:def.abs.big.succ}
\end{figure}

By (I) and (III), it suffices to compute
an incrementing CA $\acaut_\aregaut^{< \omega}$ for which:
\begin{describe}{(IV)}
\item[(IV)]
$\acaut_\aregaut^{< \omega}$ accepts
$\aletter_0 \ldots \aletter_{l - 1} \in \aalphabet^{< \omega}$ iff
there exists a sequence
$h_0 \Rightarrow \cdots h_{k - 1} \Rightarrow \emptyset$
of elements of $H_\aregaut$ such that:
\begin{itemize}
\item
$h_0$ is of the form
$\tuple{\aletter_0, \mathit{ee},
        \emptyset, \{\aloc_I\}, \locs_\dag \mapsto 0}$;
\item
for each $0 < i < k$,
the first component of $h_i$ is $\aletter_i$;
\item
$k = l$ if the second component of $h_{k - 1}$ is $\top$.
\end{itemize}
\end{describe}

$\acaut_\aregaut^{< \omega}$ is constructed so that it guesses and checks
a sequence $h_0 \Rightarrow \cdots h_{k - 1} \Rightarrow \emptyset$ as in (IV),
storing at most two consecutive members in any state.
To store an abstract set
$\tuple{\aletter, \mathit{ee},
        \locs_=, \locs_\emptyset, \sharp}$,
locations of $\acaut_\aregaut^{< \omega}$
are used for the first four components,
and $\sharp$ is stored by means of $2^{\length{\locs}} - 1$ counters
$c_{\locs_\dag}$ for $\emptyset \neq \locs_\dag \subseteq \locs$.
$\acaut_\aregaut^{< \omega}$ also has $4^{\length{\locs}}$ auxiliary counters
$c_{\locs'_{\neq}, \locs'_=}$ for $\locs'_{\neq}, \locs'_= \subseteq \locs$.
The nontrivial part of $\acaut_\aregaut^{< \omega}$ is,
given $\aletter$, $\mathit{ee}$, $\locs_=$, $\locs_\emptyset$,
and $\sharp$ which is stored by the counters $c_{\locs_\dag}$,
to guess maps $\amap_=$, $\amap_\emptyset$ and $\amap_{\locs_\dag, j}$,
and set the counters $c_{\locs_\dag}$ so that
they store $\sharp'$ as defined above.
Figure~\ref{f:comp.abs.big.succ} contains pseudo-code for that computation,
which is by $\emptyword$ transitions.
It assumes that each $c_{\locs'_{\neq}, \locs'_=}$
is zero at the beginning, and ensures that the same holds at the end.
The choices of maps are nondeterministic.
If a map cannot be chosen because a corresponding set
$\Succ{\aletter, \mathit{ee}, \mathit{uu}, \aloc}$ is empty,
the computation blocks.

\begin{narrowfig}{.6\textwidth}
\begin{tabbing}
\textbf{for all} $\emptyset \neq \locs_\dag \subseteq \locs$ \\
\{ \= \textbf{while} $c_{\locs_\dag} > 0$ \\
   \> \{ \= \textbf{choose a map}
              $\aloc \in \locs_\dag \,\mapsto\, \amap(\aloc) \in
                 \Succ{\aletter, \mathit{ee}, \bot, \aloc}$; \\
   \>    \> $\mathtt{dec}(c_{\locs_\dag})$;
            $\mathtt{inc}(c_{\bigcup_1 \amap, \bigcup_2 \amap})$ \} \}; \\
\textbf{choose a map}
  $\aloc \in \locs_= \,\mapsto\, \amap_=(\aloc) \in
     \Succ{\aletter, \mathit{ee}, \top, \aloc}$; \\
\textbf{choose a map}
  $\aloc \in \locs_\emptyset \,\mapsto\, \amap_\emptyset(\aloc) \in
     \Succ{\aletter, \mathit{ee}, \bot, \aloc}$; \\
$\locs_\ddag := \bigcup_2 \amap_= \,\cup\, \bigcup_2 \amap_\emptyset$; \\
\textbf{for all} $\locs'_{\neq}, \locs'_= \subseteq \locs$ \\
\{ \= \textbf{if} $c_{\locs'_{\neq}, \locs'_=} > 0$
      \textbf{then} $\locs_\ddag := \locs_\ddag \,\cup\, \locs'_=$ \}; \\
\textbf{if} $\locs_\ddag \neq \emptyset$
\textbf{then} $\mathtt{inc}(c_{\locs_\ddag})$; \\
\textbf{for all} $\locs'_{\neq}, \locs'_= \subseteq \locs$ \\
\{ \= \textbf{while} $c_{\locs'_{\neq}, \locs'_=} > 0$ \\
   \> \{ \= $\mathtt{dec}(c_{\locs'_{\neq}, \locs'_=})$;
            \textbf{if} $\locs'_{\neq} \neq \emptyset$
            \textbf{then} $\mathtt{inc}(c_{\locs'_{\neq}})$ \} \}
\end{tabbing}
\caption{Computing an abstract \\
         big-step successor}
\label{f:comp.abs.big.succ}
\end{narrowfig}

Suppose $\aletter_0 \ldots \aletter_{l - 1} \in \aalphabet^{< \omega}$.
If there exists a sequence
$h_0 \Rightarrow \cdots h_{k - 1} \Rightarrow \emptyset$
as in (IV), the construction of $\acaut_\aregaut^{< \omega}$
ensures that it accepts $\aletter_0 \ldots \aletter_{l - 1}$
by a run without incrementing errors.
For the reverse direction in (IV), let
$\tuple{\aletter, \mathit{ee},
        \locs_=, \locs_\emptyset, \sharp}
 \sqsubseteq
 \tuple{\aletter', \mathit{ee}',
        \locs'_=, \locs'_\emptyset, \sharp'}$
mean that
$\aletter = \aletter'$,
$\mathit{ee} = \mathit{ee}'$,
$\locs_= \subseteq \locs'_=$,
$\locs_\emptyset \subseteq \locs'_\emptyset$ and
there exists an injective
\[\iota:
  \{\tuple{\locs_\dag, j} \,:\,
    j \in \{1, \ldots, \sharp(\locs_\dag)\}\} \rightarrow
  \{\tuple{\locs'_\dag, j'} \,:\,
    j' \in \{1, \ldots, \sharp'(\locs'_\dag)\}\}\]
for which $\locs_\dag \subseteq \locs'_\dag$
whenever $\iota(\tuple{\locs_\dag, j}) = \tuple{\locs'_\dag, j'}$.
Also, let $\emptyset \sqsubseteq h'$ for all $h' \in H_\aregaut$.
If $\acaut_\aregaut^{< \omega}$ accepts $\aletter_0 \ldots \aletter_{l - 1}$
(by a run possibly with incrementing errors),
we have that there exist
$h'_0, \ldots, h'_{k - 1} \in H_\aregaut$ such that:
\begin{itemize}
\item
$h'_0$ is of the form
$\tuple{\aletter_0, \mathit{ee},
        \emptyset, \{\aloc_I\}, \locs_\dag \mapsto 0}$;
\item
for each $0 < i < k$,
the second component of $h'_{i - 1}$ is $\bot$,
the first component of $h'_i$ is $\aletter_i$, and
$h'_{i - 1} \Rightarrow h_i$ for some $h_i \sqsubseteq h'_i$;
\item
$h'_{k - 1} \Rightarrow \emptyset$, and
$k = l$ if the second component of $h'_{k - 1}$ is $\top$.
\end{itemize}
It remains to observe that $\sqsubseteq$ is transitive,
and downwards compatible with $\Rightarrow$,
i.e.\ whenever $\emptyset \neq h \sqsubseteq h'$ and $h' \Rightarrow h'_*$,
there exists $h_*$ such that $h \Rightarrow h_*$ and $h_* \sqsubseteq h'_*$.

The computation of $\acaut_\aregaut^\omega$ follows the same pattern,
except that the construction in the proof of
\cite[Theorem~5.1]{Miyano&Hayashi84} is used to replace
existence of winning strategies in two-player weak games by
existence of sequences of pairs of sets which satisfy a B\"uchi condition.
Specifically, for sets $\poss \neq \emptyset$ and $\poss'$
of states of $\aregaut$ for a data word $\adataword$ over $\aalphabet$,
and subsets $\poss_\flat$ of $\poss$ and $\poss'_\flat$ of $\poss'$
which consist only of states with odd ranks,
we write $\tuple{\poss, \poss_\flat} \Rightarrow \tuple{\poss', \poss'_\flat}$
iff there exists a map $\apos \mapsto \poss'_\apos$ on $\poss$ such that
$\apos \Rightarrow \poss'_\apos$ for each $\apos \in \poss$,
$\poss' = \bigcup_{\apos \in \poss} \poss'_\apos$,
$\poss'_\flat = \poss'_\natural$
if $\poss'_\natural \neq \emptyset$, and
$\poss'_\flat = \{\apos' \in \poss' \,:\, \rank(\apos') \mathrm{\ is\ odd}\}$
if $\poss'_\natural = \emptyset$, where
\[\poss'_\natural =
  \{\apos' \,:\, \mathrm{for\ some\ } \apos \in \poss_\flat,
                 \apos' \in \poss'_\apos \mathrm{\ and\ }
                 \rank(\apos') = \rank(\apos)\}\]
When $\poss'_\natural = \emptyset$
for some such map $\apos \mapsto \poss'_\apos$ on $\poss$,
we say that $\poss'_\flat$ is \emph{fresh}.
Instead of (I) above, we have:
\begin{describe}{(V)}
\item[(V)]
for every infinite data word $\adataword$ over $\aalphabet$,
$\aregaut$ accepts $\adataword$ iff
there exists a sequence
$\tuple{\poss_0, \poss_{0, \flat}} \Rightarrow
 \tuple{\poss_1, \poss_{1, \flat}} \Rightarrow \cdots$
of pairs of sets of states of $\aregaut$ for $\adataword$ such that
$\poss_0 = \{\tuple{0, \aloc_I, \emptyset}\}$,
$\poss_{0, \flat} = \poss_0$ if $\rank(\aloc_I)$ is odd,
$\poss_{0, \flat} = \emptyset$ if $\rank(\aloc_I)$ is even, and
either the sequence ends with $\tuple{\emptyset, \emptyset}$
or $\poss_{i, \flat}$ is fresh for infinitely many $i$.
\end{describe}

Finally, we remark that
$\acaut_\aregaut^{< \omega}$ and $\acaut_\aregaut^\omega$
are computable in polynomial space.
The pseudo-code in Figure~\ref{f:comp.abs.big.succ}
can be implemented so that at most one component of the maps
$\aloc \in \locs_\dag \,\mapsto\, \amap(\aloc) \in
   \Succ{\aletter, \mathit{ee}, \bot, \aloc}$,
$\aloc \in \locs_= \,\mapsto\, \amap_=(\aloc) \in
   \Succ{\aletter, \mathit{ee}, \top, \aloc}$ and
$\aloc \in \locs_\emptyset \,\mapsto\, \amap_\emptyset(\aloc) \in
   \Succ{\aletter, \mathit{ee}, \bot, \aloc}$
is stored in any state (by means of its location).
The definition in Figure~\ref{f:def.abs.big.succ}
provides a nondeterministic algorithm which,
given
$\aletter \in \aalphabet$,
$\mathit{ee}, \mathit{uu} \in \{\top, \bot\}$,
$\aloc \in \locs$ and
$\locs'_{\neq}, \locs'_= \subseteq \locs$,
checks whether 
$\tuple{\locs'_{\neq}, \locs'_=} \in
 \Succ{\aletter, \mathit{ee}, \mathit{uu}, \aloc}$
in space polynomial in the size of $\aregaut$.
\end{proof}

\section{Lower Complexity Bounds}
\label{section-lower-bounds}

To warm up, we show that the upper bounds in Theorem~\ref{th:pspace}
are tight already for deterministic automata,
and in the case of the \textsc{NLogSpace}-memberships,
already with no registers.

\begin{theorem}
\label{th:pspace-hard}
The following hold over finite and over infinite data words:
\begin{itemize}
\item[(a)]
nonemptiness for 1DRA is \textsc{PSpace}-hard;
\item[(b)]
nonemptiness for 1DRA$_0$ is \textsc{NLogSpace}-hard.
\end{itemize}
\end{theorem}

\begin{proof}
Part~(b) is an immediate consequence of
\textsc{NLogSpace}-hardness of nonemptiness for classical DFA.

For (a), we reduce from the halting problem for Turing machines
with binary alphabets and linearly bounded tapes.
Precisely, we consider Turing machines
$\amachine = \tuple{\locs, \aloc_I, \delta}$ such that
$\locs$ is a finite set of locations,
$\aloc_I$ is the initial location, and
$\delta: \locs \times \{0, 1\} \rightarrow
         \locs \times \{0, 1\} \times \{-1, 1\}$
is a transition function.
A state of $\amachine$ is a triple $\tuple{\aloc, i, \aword}$ where
$\aloc \in \locs$ is the machine location,
$0 \leq i < \length{\amachine}$ is the head position, and
$\aword \in \{0, 1\}^{\length{\amachine}}$ is the tape contents.
Let $\delta(\aloc, \aword(i)) = \tuple{\aloc', b, j}$.
If $0 \leq i + j < \length{\amachine}$,
the state $\tuple{\aloc', i + j, \aword[i \mapsto b]}$
is the unique successor of $\tuple{\aloc, i, \aword}$.
Otherwise, $\tuple{\aloc, i, \aword}$ has no successors.
The following problem is \textsc{PSpace}-hard:
given $\amachine$ as above, to decide whether the computation
from the initial state $\tuple{\aloc_I, 0, 0 0 \ldots 0}$
reaches a state with no successor.

We encode a computation
$\tuple{\aloc_0, i_0, \aword_0}
 \tuple{\aloc_1, i_1, \aword_1}
 \cdots$
of $\amachine$ by the following data word
over the alphabet $\locs \uplus \{\mbox{-}\}$.
Its underlying word is
\[\mbox{-} \, \mbox{-} \,
  \aletter^0_0 \, \aletter^0_1 \,\ldots\, \aletter^0_{\length{\amachine} - 1} \,
  \aletter^1_0 \, \aletter^1_1 \,\ldots\, \aletter^1_{\length{\amachine} - 1} \,
  \ldots\]
where $\aletter^k_l = \aloc_k$ if $l = i_k$,
and $\aletter^k_l = \mbox{-}$ otherwise.
There are two equivalence classes:
$0 \not\sim 1$,
$2 + \length{\amachine} \times k + l \,\sim\, 0$ if $\aword_k(l) = 0$, and
$2 + \length{\amachine} \times k + l \,\sim\, 1$ if $\aword_k(l) = 1$.

It is straightforward to construct,
in space logarithmic in $\length{\amachine}$,
an automaton $\aregaut_\amachine$ in 1DRA
with alphabet $\locs \uplus \{\mbox{-}\}$
which accepts a data word iff
it has a prefix that encodes a computation of $\amachine$
from the initial state to a state with no successor.
$\aregaut_\amachine$ has $2 + \length{\amachine}$ registers
$r_0$, $r_1$, and $r'_l$ for $0 \leq l < \length{\amachine}$.
It stores $[0]_\sim$ into $r_0$ and $[1]_\sim$ into $r_1$,
and checks that $0 \not\sim 1$ and the initial state is encoded correctly.
Whenever $\aregaut_\amachine$ moves to a word position
$2 + \length{\amachine} \times (k + 1)$,
it has kept $\aloc_k$ and $i_k$ using its location
and has stored $[2 + \length{\amachine} \times k + l]_\sim$ in $r'_l$
for each $0 \leq l < \length{\amachine}$.
If $\tuple{\aloc_k, i_k, \aword_k}$ has no successor,
$\aregaut_\amachine$ accepts.
Otherwise, it checks that positions
$2 + \length{\amachine} \times (k + 1)$, \ldots,
$2 + \length{\amachine} \times (k + 2) - 1$
encode the successor state,
simultaneously updates $r'_0$, $r'_1$, \ldots, $r'_{\length{\amachine} - 1}$,
and repeats.
\end{proof}

Satisfiability for LTL$^\downarrow_1(\nextt, \until)$ and
nonemptiness for 1ARA$_1$
were shown in Theorem~\ref{th:register2counter} to be
decidable over finite data words,
and in $\Pi^0_1$ over infinite data words.
We now establish their
non-primitive recursiveness in the finitary case,
and $\Pi^0_1$-hardness in the infinitary case.
In fact, we have those lower bounds even for
the unary logical fragment and universal automata.

\begin{theorem}
\label{th:counter2}
Satisfiability for LTL$^\downarrow_1(\nextt, \sometimes)$ and
nonemptiness for 1URA$_1$ are
not primitive recursive over finite data words,
and $\Pi^0_1$-hard over infinite data words.
\end{theorem}

\begin{proof}
By Theorem~\ref{th:CA} (b), it suffices to show that,
given an incrementing CA
$\acaut = \tuple{\aalphabet, \locs, \aloc_I, n, \delta, F}$,
sentences $\aformula_\acaut^{< \omega}$ and $\aformula_\acaut^\omega$
of LTL$^\downarrow_1(\nextt, \sometimes)$ and
automata $\aregaut_\acaut^{< \omega}$ and $\aregaut_\acaut^\omega$
in 1URA$_1$ are computable in logarithmic space,
such that their alphabet is
$\widehat{\aalphabet} =
 \locs \times (\aalphabet \cup \{\emptyword\}) \times L \times \locs$
where $L = \{\mathtt{inc, dec, ifz}\} \times \{1, \ldots, n\}$, and
\[\mathrm{L}^\alpha(\acaut)
= \{\overline{\adataword} \,:\,
    \adataword \in \mathrm{L}^\alpha(\aformula_\acaut^\alpha)\}
= \{\overline{\adataword} \,:\,
    \adataword \in \mathrm{L}^\alpha(\aregaut_\acaut^\alpha)\}\]
for $\alpha \in \{{<} \omega, \omega\}$,
where $\overline{\adataword} = \aword_0 \aword_1 \ldots$ if
$\mathrm{str}(\adataword) =
 \tuple{\aloc_0, \aword_0, l_0, \aloc'_0}
 \tuple{\aloc_1, \aword_1, l_1, \aloc'_1} \cdots$.

To ensure that a data word over $\widehat{\aalphabet}$
encodes a run of $\acaut$, we constrain its equivalence relation.
Firstly, there must not be two $\tuple{\mathtt{inc}, c}$ transitions
or two $\tuple{\mathtt{dec}, c}$ transitions
(with the same $c$) in the same class.
For an $\tuple{\mathtt{ifz}, c}$ transition to be correct,
whenever it is preceded by $\tuple{\mathtt{inc}, c}$,
there must be an intermediate $\tuple{\mathtt{dec}, c}$ in the same class.
Incrementing errors may occur because a $\tuple{\mathtt{dec}, c}$ transition
may be preceded by no $\tuple{\mathtt{inc}, c}$ in the same class.
Such a $\tuple{\mathtt{dec}, c}$ transition corresponds to
a faulty decrement which leaves $c$ unchanged.
It is easy to check that, for every run of $\acaut$,
there exists a run which differs at most in counter values
and whose only incrementing errors are such faulty decrements.

More precisely, $\acaut$ accepts a finite word $\aword$ over $\aalphabet$ iff
$\aword = \overline{\adataword}$ for some finite data word
$\adataword$ over $\widehat{\aalphabet}$ satisfying the following, where
$\mathrm{str}(\adataword) =
 \tuple{\aloc_0, \aword_0, l_0, \aloc'_0}
 \tuple{\aloc_1, \aword_1, l_1, \aloc'_1} \cdots$:
\begin{itemize}
\item[(1)]
for each $i$, $\tuple{\aloc_i, \aword_i, l_i, \aloc'_i} \in \delta$;
\item[(2)]
$\aloc_0 = \aloc_I$, and for each $i > 0$,
$\aloc'_{i - 1} = \aloc_i$;
\item[(3)]
for the maximum $i$, $\aloc'_i \in F$;
\item[(4)]
there are no $c$ and $i < j$ such that
$l_i = l_j = \tuple{\mathtt{inc}, c}$ and
$i \sim^\adataword j$;
\item[(5)]
there are no $c$ and $i < j$ such that
$l_i = l_j = \tuple{\mathtt{dec}, c}$ and
$i \sim^\adataword j$;
\item[(6)]
for all $c$ and $i$ such that $l_i = \tuple{\mathtt{inc}, c}$,
it is not the case that,
there is $j > i$ with $l_j = \tuple{\mathtt{ifz}, c}$
but there is no $k > i$ with
$l_k = \tuple{\mathtt{dec}, c}$ and $i \sim^\adataword k$;
\item[(7)]
there are no $c$ and $i < j < k$ such that
$l_i = \tuple{\mathtt{inc}, c}$,
$l_j = \tuple{\mathtt{ifz}, c}$,
$l_k = \tuple{\mathtt{dec}, c}$ and
$i \sim^\adataword k$.
\end{itemize}

$\aformula_\acaut^{< \omega}$ is constructed to express
the conjunction of (1)--(7).
(1)--(3) are straightforward.
Among (4)--(7), the most interesting is (7),
and the rest can be expressed similarly.
Observe how (6) and (7) were formulated to avoid using the $\until$ operator.
The following sentence expresses (7):
\[\begin{array}{r}
\neg \bigvee_{c = 1}^n \sometimes \bigg(
  (\bigvee_{\aloc, \aword, \aloc'}
   \tuple{\aloc, \aword, \tuple{\mathtt{inc}, c}, \aloc'}) \wedge
  {\downarrow_1} \nextt \sometimes \Big(
    (\bigvee_{\aloc, \aword, \aloc'}
     \tuple{\aloc, \aword, \tuple{\mathtt{ifz}, c}, \aloc'}) \wedge \\
    \nextt \sometimes \big(
      (\bigvee_{\aloc, \aword, \aloc'}
       \tuple{\aloc, \aword, \tuple{\mathtt{dec}, c}, \aloc'}) \wedge
      {\uparrow_1}
    \big)
  \Big)
\bigg)
\end{array}\]

For $\aregaut_\acaut^{< \omega}$, it is sufficient by Theorem~\ref{th:closure},
for each of (1)--(7), to compute in logarithmic space an automaton in 1NRA$_1$
which accepts a finite data word over $\widehat{\aalphabet}$
iff it fails the condition.
In fact, (6) and (7) can be treated together by checking that
some $\tuple{\mathtt{inc}, c}$ instruction is followed by
no occurence of $\tuple{\mathtt{dec}, c}$ with the same datum
until $\tuple{\mathtt{ifz}, c}$ occurs,
and this automaton is the most interesting.
It is shown in Figure~\ref{f:wrong.ifz}, where
$\tuple{\aloc_1, \aword_1, \aloc'_1}, \ldots,
 \tuple{\aloc_K, \aword_K, \aloc'_K}$ enumerates
$\locs \times (\aalphabet \cup \{\emptyword\}) \times \locs$.

\begin{figure}
\setlength{\unitlength}{1.75em}
\begin{center}
\begin{picture}(24,9)(-.5,.5)
\gasset{Nw=.6,Nh=.6,Nmr=.3,Nadjust=wh,ExtNL=n,NLdist=0}
\node[Nmarks=i,iangle=180](i)(1,9){$\vee$}
\node(iX)(1,7){$\nextt$}
\drawedge[curvedepth=.5](i,iX){}
\drawedge[curvedepth=.5](iX,i){}
\node(c)(3,9){$\bigvee_{c = 1}^n$}
\drawedge(i,c){}
\node(i1)(7,9){$\tuple{\aloc_1, \aword_1, \tuple{\mathtt{inc}, c}, \aloc'_1}$}
\drawedge(c,i1){}
\node[Nframe=n](id)(7,7.25){$\vdots$}
\drawedge(i1,id){n}
\node(iK)(7,6){$\tuple{\aloc_K, \aword_K, \tuple{\mathtt{inc}, c}, \aloc'_K}$}
\drawedge[ELpos=67](iK,iX){n}
\node(f)(11,9){$\downarrow_1$}
\drawedge[ELpos=75](i1,f){y}
\drawedge[ELside=r](iK,f){y}
\node(fX)(13,9){$\nextt$}
\drawedge(f,fX){}
\node(d1)(17,9){$\tuple{\aloc_1, \aword_1, \tuple{\mathtt{dec}, c}, \aloc'_1}$}
\drawedge(fX,d1){}
\node[Nframe=n](dd)(17,7.25){$\vdots$}
\drawedge(d1,dd){n}
\node(dK)(17,6){$\tuple{\aloc_K, \aword_K, \tuple{\mathtt{dec}, c}, \aloc'_K}$}
\node(u)(21,9){$\uparrow_1$}
\drawedge[ELpos=75](d1,u){y}
\drawedge[ELside=r](dK,u){y}
\node(b)(23,9){$\bot$}
\drawedge(u,b){y}
\drawedge[ELside=r,ELpos=67](u,dd){n}
\node(z1)(17,4){$\tuple{\aloc_1, \aword_1, \tuple{\mathtt{ifz}, c}, \aloc'_1}$}
\drawedge(dK,z1){n}
\node[Nframe=n](zd)(17,2.25){$\vdots$}
\drawedge(z1,zd){n}
\node(zK)(17,1){$\tuple{\aloc_K, \aword_K, \tuple{\mathtt{ifz}, c}, \aloc'_K}$}
\node(t)(21,4){$\top$}
\drawedge[ELpos=75](z1,t){y}
\drawedge[ELside=r](zK,t){y}
\drawedge[curvedepth=2](zK,fX){n}
\end{picture}
\end{center}
\caption{Recognising a wrong zero test}
\label{f:wrong.ifz}
\end{figure}
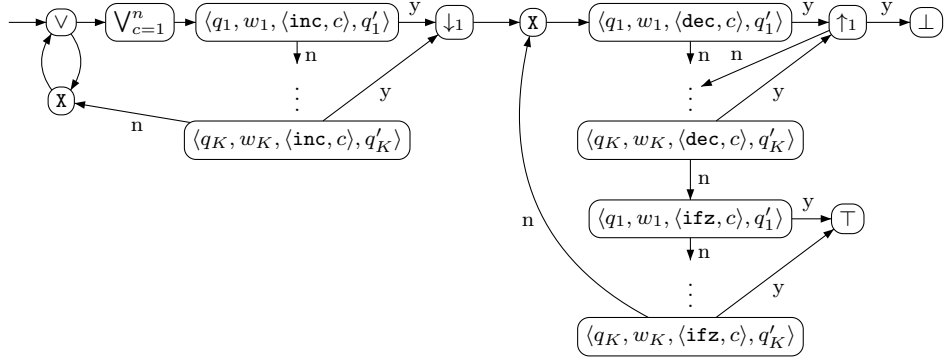

In the infinitary case, we replace (3) by:
\begin{itemize}
\item[(3')]
for infinitely many $i$, $\aloc_i \in F$.
\end{itemize}
A sentence of LTL$^\downarrow_1(\nextt, \sometimes)$ which expresses (3') is
$\always \sometimes
 \bigvee_{\aloc \in F, \aword, l, \aloc'}
 \tuple{\aloc, \aword, l, \aloc'}$.
To express the negation of (3') in 1NRA$_1$,
the automaton guesses $i$ and checks that
$\aloc_j \notin F$ for each $j \geq i$.
\end{proof}

From the proofs of Theorems \ref{th:register2counter} and \ref{th:counter2},
and by observing that incrementing CA are closed under homomorphisms,
we have the following characterisation of languages of incrementing CA
in terms of languages of future-time LTL with $1$ register.
Remarkably, it is not affected by restricting to the unary logical fragment.

\begin{corollary}
\label{cor:languages}
For both $\alpha \in \{{<} \omega, \omega\}$
and every finite alphabet $\aalphabet$, we have:
\[\begin{array}{c}
  \{\mathrm{L}^\alpha(\acaut) \,:\,
    \acaut \mathrm{\ is\ an\ incrementing\ CA\
      with\ alphabet\ } \aalphabet\} \\[.5ex]
= \{\{\amap(\mathrm{str}(\adataword)) \in \aalphabet^\alpha \,:\,
      \adataword \in \mathrm{L}^\alpha(\aformula)\} \::\\
    \aalphabet' \stackrel{\amap}{\rightarrow} \aalphabet \cup \{\emptyword\},\
    \aformula \mathrm{\ is\ a\ sentence\ of\ LTL}
      ^\downarrow_1(\nextt, \sometimes)
      \mathrm{\ with\ alphabet\ } \aalphabet'\} \\[.5ex]
= \{\{\amap(\mathrm{str}(\adataword)) \in \aalphabet^\alpha \,:\,
      \adataword \in \mathrm{L}^\alpha(\aformula)\} \::\\
    \aalphabet' \stackrel{\amap}{\rightarrow} \aalphabet \cup \{\emptyword\},\
    \aformula \mathrm{\ is\ a\ sentence\ of\ LTL}
      ^\downarrow_1(\nextt, \until)
      \mathrm{\ with\ alphabet\ } \aalphabet'\}
\end{array}\]
\end{corollary}

Our final result shows that
the problems in Theorem~\ref{th:register2counter} become
$\Sigma^0_1$-hard in the finitary case and
$\Sigma^1_1$-hard in the infinitary case
as soon as any of $1$ more register,
the $\pastsometimes$ temporal operator or
backward automaton moves are added,
even after restricting to the unary logical fragment and universal automata.
The result should also be compared with Theorem~\ref{th:basic-upper-bounds}.

The theorem below improves
\cite[Corollary~1]{French03} and
\cite[Theorem~3]{Demri&Lazic&Nowak07},
which showed $\Sigma^1_1$-hardness of
the infinitary satisfiability problems for
LTL$^\downarrow_2(\nextt, \nextt^{-1}, \sometimes, \pastsometimes)$ and
LTL$^\downarrow_2(\nextt, \until)$.
Also, recalling Theorem~\ref{th:closure},
it implies \cite[Theorem~5.1]{Neven&Schwentick&Vianu04}
where finitary nonuniversality for 1NRA was shown undecidable.
Undecidability of finitary nonemptiness for 2DRA$_1$
was shown in \cite[Section~7.3]{David04}, using a different encoding.

%Include \Sigma^1_1-hardness of nonemptiness for B\"uchi 2DRA$_1(\sim)$
%by an encoding like in the \Sigma^1_1-hardness result in our TIME paper?

%Include \Sigma^0_2-hardness of
%infinitary nonemptiness for 2DRA$_1(\sim)$
%by revising the construction for B\"uchi 2DRA$_1(\sim)$?
%If that does not work, \Sigma^0_2-hardness of
%infinitary nonemptiness for 2DRA$_2(\sim)$
%by revising the construction for B\"uchi 2DRA$_2(\sim)$
%(commented out in proof-last-theorem)?
%We thought that the following is \Sigma^0_2-complete:
%given a nondeterministic Minsky machine and location q_1,
%does there exist an infinite computation which visits q_1 exactly once?

\begin{theorem}
\label{theorem-the-last}
Over finite (resp., infinite) data words, we have that
satisfiability for LTL$^\downarrow_1(\nextt, \sometimes, \pastsometimes)$,
nonemptiness for 2DRA$_1$ (resp., 2URA$_1$),
satisfiability for LTL$^\downarrow_2(\nextt, \sometimes)$ and
nonemptiness for 1URA$_2$ are
$\Sigma^0_1$-hard (resp., $\Sigma^1_1$-hard).
\end{theorem}

\begin{proof}
By Theorem~\ref{th:CA} (a), it is sufficient to show that,
given a Minsky CA
$\acaut = \tuple{\aalphabet, \locs, \aloc_I, n, \delta, F}$,
sentences $\aformula_\acaut^{< \omega}$ and $\aformula_\acaut^\omega$
of LTL$^\downarrow_1(\nextt, \sometimes, \pastsometimes)$,
automata $\aregaut_\acaut^{< \omega}$ in 2DRA$_1$
and $\aregaut_\acaut^\omega$ in 2URA$_1$,
sentences $\aformulabis_\acaut^{< \omega}$ and $\aformulabis_\acaut^\omega$
of LTL$^\downarrow_2(\nextt, \sometimes)$ and
automata $\aregautbis_\acaut^{< \omega}$ and $\aregautbis_\acaut^\omega$
in 1URA$_2$ are computable in logarithmic space,
such that their alphabet is
$\widehat{\aalphabet} =
 \locs \times (\aalphabet \cup \{\emptyword\}) \times L \times \locs$
where $L = \{\mathtt{inc, dec, ifz}\} \times \{1, \ldots, n\}$, and
\[\begin{array}{r}
  \mathrm{L}^\alpha(\acaut)
= \{\overline{\adataword} \,:\,
    \adataword \in \mathrm{L}^\alpha(\aformula_\acaut^\alpha)\}
= \{\overline{\adataword} \,:\,
    \adataword \in \mathrm{L}^\alpha(\aregaut_\acaut^\alpha)\} \\
= \{\overline{\adataword} \,:\,
    \adataword \in \mathrm{L}^\alpha(\aformulabis_\acaut^\alpha)\}
= \{\overline{\adataword} \,:\,
    \adataword \in \mathrm{L}^\alpha(\aregautbis_\acaut^\alpha)\}
\end{array}\]
for $\alpha \in \{{<} \omega, \omega\}$,
where $\overline{\adataword} = \aword_0 \aword_1 \ldots$ if
$\mathrm{str}(\adataword) =
 \tuple{\aloc_0, \aword_0, l_0, \aloc'_0}
 \tuple{\aloc_1, \aword_1, l_1, \aloc'_1} \cdots$.

To ensure that a data word over $\widehat{\aalphabet}$
corresponds to a run of $\acaut$, we constrain its equivalence relation
as was done for incrementing CA in the proof of Theorem~\ref{th:counter2},
and in addition require that each $\tuple{\mathtt{dec}, c}$ transition
be preceded by some $\tuple{\mathtt{inc}, c}$ in the same class,
which eliminates the possibility of faulty decrements.

More precisely, $\acaut$ accepts a
finite (resp., infinite) word $\aword$ over $\aalphabet$
iff $\aword = \overline{\adataword}$ for some
finite (resp., infinite) data word $\adataword$ over $\widehat{\aalphabet}$
which satisfies (1)--(7) (resp., (1), (2), (3') and (4)--(7))
given in the proof of Theorem~\ref{th:counter2}, and
\begin{itemize}
\item[(8)]
whenever $l_i = \tuple{\mathtt{dec}, c}$,
there exists $j < i$ such that
$l_j = \tuple{\mathtt{inc}, c}$ and $i \sim^\adataword j$,
\end{itemize}
where
$\mathrm{str}(\adataword) =
 \tuple{\aloc_0, \aword_0, l_0, \aloc'_0}
 \tuple{\aloc_1, \aword_1, l_1, \aloc'_1} \cdots$.

To compute $\aformula_\acaut^{< \omega}$ and $\aformula_\acaut^\omega$,
(8) is expressible in LTL$^\downarrow_1(\nextt, \sometimes, \pastsometimes)$ as:
\[\textstyle{\bigwedge}_{c = 1}^n \always \Big(
    (\textstyle{\bigvee}_{\aloc, \aword, \aloc'}
     \tuple{\aloc, \aword, \tuple{\mathtt{dec}, c}, \aloc'}) \Rightarrow
    {\downarrow_1} \pastsometimes \big(
      (\textstyle{\bigvee}_{\aloc, \aword, \aloc'}
       \tuple{\aloc, \aword, \tuple{\mathtt{inc}, c}, \aloc'}) \wedge
      {\uparrow_1}
    \big)
  \Big)\]

$\aregaut_\acaut^{< \omega}$ is constructed to check (1)--(8) as follows:
\begin{itemize}
\item
if the current transition, and the previous one (if any),
fail (1) or (2), $\aregaut_\acaut^{< \omega}$ rejects;
\item
%[$\dag$]
if the current transition fails (3), $\aregaut_\acaut^{< \omega}$ rejects;
\item
%[$\ddag$]
if the current instruction is $\tuple{\mathtt{inc}, c}$,
$\aregaut_\acaut^{< \omega}$ stores the current class in the register,
and then scans $\adataword$ forwards and rejects if it finds
an $\tuple{\mathtt{inc}, c}$ in the same class,
or an $\tuple{\mathtt{ifz}, c}$ before
a $\tuple{\mathtt{dec}, c}$ in the same class,
but otherwise returns;
\item
if the current instruction is $\tuple{\mathtt{dec}, c}$,
$\aregaut_\acaut^{< \omega}$ stores the current class in the register,
and then scans $\adataword$ backwards and rejects if it finds
a $\tuple{\mathtt{dec}, c}$ in the same class,
or no $\tuple{\mathtt{inc}, c}$ in the same class,
but otherwise returns;
\item
if there is a next transition,
$\aregaut_\acaut^{< \omega}$ repeats the above for it,
but otherwise accepts.
\end{itemize}

$\aregaut_\acaut^\omega$ is constructed by extending the construction of
$\aregaut_\acaut^\omega$ in the proof of Theorem~\ref{th:counter2}
by expressing the negation of (8) by an automaton in 2NRA$_1$:
it guesses a position with a $\tuple{\mathtt{dec}, c}$ instruction,
and checks that it is not preceded by
an $\tuple{\mathtt{inc}, c}$ in the same class.

For the logical fragments and automata classes with $2$ registers,
we use a different encoding of runs of $\acaut$ by data words,
similar to the modelling in \cite[Section~4]{Lisitsa&Potapov05}
of Minsky machines by systems of pebbles.
Let
$\widetilde{\aalphabet} =
 \widehat{\aalphabet} \cup
 (\{\mathtt{hi}, \mathtt{lo}\} \times \{1, \ldots, n\})$.
For a data word $\adataword$ over $\widetilde{\aalphabet}$, let
$\overline{\adataword} =
 \overline{\adataword \upharpoonright \widehat{\aalphabet}}$.

Each transition $\tuple{\aloc, \aword, l, \aloc'}$ in a run of $\acaut$
is encoded by a block whose sequence of letters is
$\tuple{\mathtt{hi}, 1} \tuple{\mathtt{lo}, 1} \cdots
 \tuple{\mathtt{hi}, n} \tuple{\mathtt{lo}, n}
 \tuple{\aloc, \aword, l, \aloc'}$.
For each counter $c$,
two occurences of $\tuple{\mathtt{hi}, c}$ are in the same class
iff there is no occurence of $\tuple{\mathtt{inc}, c}$ between them,
which gives a sequence of classes $D_0$, $D_1$, \ldots
Each occurence of $\tuple{\mathtt{lo}, c}$ is in some class $D_i$.
If prior to a transition of $\acaut$, a counter $c$ has value $m$,
that is encoded by occurences of
$\tuple{\mathtt{lo}, c}$ and $\tuple{\mathtt{hi}, c}$
in the corresponding block being in some classes $D_i$ and $D_{i + m}$.

More precisely, we have that $\acaut$ accepts a
finite (resp., infinite) word $\aword$ over $\aalphabet$
iff $\aword = \overline{\adataword}$ for some
finite (resp., infinite) data word $\adataword$ over $\widetilde{\aalphabet}$
which satisfies:
\begin{itemize}
\item[(i)]
$\mathrm{str}(\adataword)$ is a sequence of blocks
$\tuple{\mathtt{hi}, 1} \tuple{\mathtt{lo}, 1} \cdots
 \tuple{\mathtt{hi}, n} \tuple{\mathtt{lo}, n}
 \tuple{\aloc, \aword, l, \aloc'}$;
\item[(ii)]
each $\tuple{\aloc, \aword, l, \aloc'}$ is in $\delta$;
\item[(iii)]
for the first $\tuple{\aloc, \aword, l, \aloc'}$, $\aloc = \aloc_I$,
and for each
$\tuple{\aloc, \aword, l, \aloc'}$ and
$\tuple{\aloc'', \aword', l', \aloc'''}$
which are consecutive,
$\aloc' = \aloc''$;
\item[(iv)]
for the last $\tuple{\aloc, \aword, l, \aloc'}$, $\aloc' \in F$
(resp., infinitely often $\aloc \in F$);
\item[(v)]
in the inital block, for each $c$,
$\tuple{\mathtt{hi}, c}$ and $\tuple{\mathtt{lo}, c}$
are in the same class;
\item[(vi)]
in each block immediately after an $\tuple{\mathtt{inc}, c}$ one,
$\tuple{\mathtt{hi}, c}$ is not in the same class as
any preceding $\tuple{\mathtt{hi}, c}$, and
$\tuple{\mathtt{lo}, c}$ is in the same class as
the previous $\tuple{\mathtt{lo}, c}$;
\item[(vii)]
in each $\tuple{\mathtt{dec}, c}$ block,
$\tuple{\mathtt{hi}, c}$ and $\tuple{\mathtt{lo}, c}$
are not in the same class;
\item[(viii)]
in each block immediately after a $\tuple{\mathtt{dec}, c}$ block $B$,
$\tuple{\mathtt{hi}, c}$ is in the same class as
the previous $\tuple{\mathtt{hi}, c}$, and
$\tuple{\mathtt{lo}, c}$ is in the same class as
$\tuple{\mathtt{hi}, c}$ in the block immediately after
the last block containing $\tuple{\mathtt{hi}, c}$
which is in the same class as $\tuple{\mathtt{lo}, c}$ in $B$;
\item[(ix)]
in each $\tuple{\mathtt{ifz}, c}$ block,
$\tuple{\mathtt{hi}, c}$ and $\tuple{\mathtt{lo}, c}$
are in the same class.
\end{itemize}

For $\aformulabis_\acaut^{< \omega}$ and $\aformulabis_\acaut^\omega$,
each of (i)--(ix) is expressed in LTL$^\downarrow_2(\nextt, \sometimes)$.
In fact, (viii) naturally splits into two halves,
and the second half is the most involved among (i)--(ix):
\[\begin{array}{c}
\always \bigwedge_{c = 1}^n \bigg(
  \tuple{\mathtt{hi}, c} \Rightarrow
  {\downarrow_1} \nextt^{2 n + 1} \Big(
    \neg {\uparrow_1} \Rightarrow \\
    {\downarrow_2} \always \big(
      \tuple{\mathtt{lo}, c} \wedge {\uparrow_1} \wedge
        (\nextt^{2 (n - c) + 1}
         \bigvee_{\aloc, \aword, \aloc'}
         \tuple{\aloc, \aword, \tuple{\mathtt{dec}, c}, \aloc'}) \Rightarrow
      \nextt^{2 n + 1} {\uparrow_2}
    \big)
  \Big)
\bigg)
\end{array}\]

It remains by Theorem~\ref{th:closure}, for each of (i)--(ix),
to compute in logarithmic space an automaton in 1NRA$_2$ which accepts a
finite (resp., infinite) data word $\adataword$ over $\widetilde{\aalphabet}$
iff it fails the condition.
For the second half of (viii), which is again the most involved, the automaton
guesses a position with a $\tuple{\mathtt{hi}, c}$ letter,
checks that the position $2 n + 1$ steps forwards
(which is the next occurence of $\tuple{\mathtt{hi}, c}$)
is not in the same class,
guesses a subsequent position with the $\tuple{\mathtt{lo}, c}$ letter
which is in the same class as the first $\tuple{\mathtt{hi}, c}$ position
and whose block ends with the $\tuple{\mathtt{dec}, c}$ instruction, and
checks that the position $2 n + 1$ steps forwars
(which is the next occurence of $\tuple{\mathtt{lo}, c}$)
is not in the same class as the second $\tuple{\mathtt{hi}, c}$ position.
\end{proof}

\section{Conclusion}
\label{section-conclusion} 

By Theorems \ref{th:basic-upper-bounds}, \ref{th:register2counter},
\ref{th:counter2} and \ref{theorem-the-last}, we have the results
on complexity of satisfiability shown in Figure~\ref{f:compl.sat},
where `R, not PR' means decidable and not primitive recursive.
The entries not in bold follow from
\cite[Corollary~1]{French03} and
\cite[Theorem~3]{Demri&Lazic&Nowak07}.

\begin{figure}
\begin{center}
\begin{tabular}{r|c|c|c|c|}
& \multicolumn{2}{c|}{finite data words}
& \multicolumn{2}{c|}{infinite data words}
\\
registers
& $1$ & $2$
& $1$ & $2$
\\ \hline
$\nextt, \sometimes$
& \textbf{R, not PR}
& $\mathbf{\Sigma^0_1}$\textbf{-complete}
& $\mathbf{\Pi^0_1}$\textbf{-complete}
& $\mathbf{\Sigma^1_1}$\textbf{-complete}
\\ \hline
$\nextt, \until$
& \textbf{R, not PR}
& $\Sigma^0_1$-complete
& $\mathbf{\Pi^0_1}$\textbf{-complete}
& $\Sigma^1_1$-complete
\\ \hline
$\nextt, \sometimes, \pastsometimes$
& $\mathbf{\Sigma^0_1}$\textbf{-complete}
& $\mathbf{\Sigma^0_1}$\textbf{-complete}
& $\mathbf{\Sigma^1_1}$\textbf{-complete}
& $\mathbf{\Sigma^1_1}$\textbf{-complete}
\\ \hline
\end{tabular}
\end{center}
\caption{Complexity of satisfiability for fragments of
         LTL with the freeze quantifier}
\label{f:compl.sat}
\end{figure}

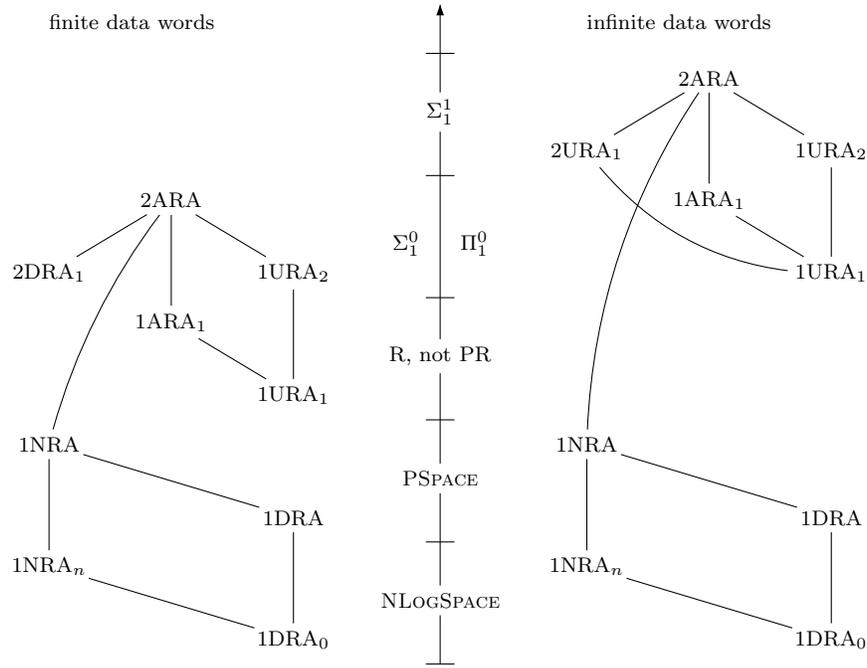
\begin{figure}
\setlength{\unitlength}{0.65mm}
\begin{center}
\begin{picture}(160,135)
\put(0,130){finite data words}

\node[Nframe=n,Nadjust=wh](2af)(25,95){2ARA}
\node[Nframe=n,Nadjust=wh](2df1)(0,80){2DRA$_1$}
\node[Nframe=n,Nadjust=wh](1u2f)(50,80){1URA$_2$}

\node[Nframe=n,Nadjust=wh](1a1f)(25,70){1ARA$_1$}
\node[Nframe=n,Nadjust=wh](1u1f)(50,55){1URA$_1$}

\node[Nframe=n,Nadjust=wh](1nf)(0,45){1NRA}
\node[Nframe=n,Nadjust=wh](1df)(50,30){1DRA}

\node[Nframe=n,Nadjust=wh](1nfk)(0,20){1NRA$_n$}
\node[Nframe=n,Nadjust=wh](1dfk)(50,5){1DRA$_0$}

\drawedge[AHnb=0](2af,2df1){}
\drawedge[AHnb=0](2af,1u2f){}
\drawedge[AHnb=0](2af,1a1f){}
\drawedge[AHnb=0](1u2f,1u1f){}
\drawedge[AHnb=0,curvedepth=-3](2af,1nf){}
\drawedge[AHnb=0](1a1f,1u1f){}
\drawedge[AHnb=0](1nf,1df){}
\drawedge[AHnb=0](1nfk,1dfk){}
\drawedge[AHnb=0](1nf,1nfk){}
\drawedge[AHnb=0](1df,1dfk){}

\put(110,130){infinite data words}

\node[Nframe=n,Nadjust=wh](2ai)(135,120){2ARA}
\node[Nframe=n,Nadjust=wh](2ui1)(110,105){2URA$_1$}
\node[Nframe=n,Nadjust=wh](1ui2)(160,105){1URA$_2$}

\node[Nframe=n,Nadjust=wh](1ai1)(135,95){1ARA$_1$}
\node[Nframe=n,Nadjust=wh](1ui1)(160,80){1URA$_1$}

\node[Nframe=n,Nadjust=wh](1ni)(110,45){1NRA}
\node[Nframe=n,Nadjust=wh](1di)(160,30){1DRA}

\node[Nframe=n,Nadjust=wh](1nik)(110,20){1NRA$_n$}
\node[Nframe=n,Nadjust=wh](1dik)(160,5){1DRA$_0$}

\drawedge[AHnb=0](2ai,2ui1){}
\drawedge[AHnb=0](2ai,1ui2){}
\drawedge[AHnb=0](2ai,1ai1){}
\drawedge[AHnb=0](1ui2,1ui1){}
\drawedge[AHnb=0](1ai1,1ui1){}
\drawedge[AHnb=0,curvedepth=-6](2ai,1ni){}
\drawedge[AHnb=0](1ni,1di){}
\drawedge[AHnb=0](1nik,1dik){}
\drawedge[AHnb=0](1ni,1nik){}
\drawedge[AHnb=0](1di,1dik){}
\drawedge[AHnb=0,curvedepth=-7](2ui1,1ui1){}

\node[Nframe=n,Nw=0,Nh=0](node1)(77,125){}
\node[Nframe=n,Nw=0,Nh=0](node2)(77,100){}
\node[Nframe=n,Nw=0,Nh=0](node3)(77,75){}
\node[Nframe=n,Nw=0,Nh=0](node4)(77,50){}
\node[Nframe=n,Nw=0,Nh=0](node5)(77,25){}
\node[Nframe=n,Nw=0,Nh=0](node6)(77,0){}

\node[Nframe=n,Nw=0,Nh=0](node7)(80,135){}
\node[Nframe=n,Nadjust=wh](node8)(80,113){$\Sigma_1^1$}
\node[Nframe=n,Nadjust=wh](node8l)(73,86){$\Sigma_1^0$}
\node[Nframe=n,Nadjust=wh](node8r)(87,86){$\Pi_1^0$}
\node[Nframe=n,Nadjust=wh](node9)(80,63){R, not PR}
\node[Nframe=n,Nadjust=wh](node10)(80,38){\textsc{PSpace}}
\node[Nframe=n,Nadjust=wh](node11)(80,13){\textsc{NLogSpace}}
\node[Nframe=n,Nw=0,Nh=0](node12)(80,0){}

\node[Nframe=n,Nw=0,Nh=0](node13)(83,125){}
\node[Nframe=n,Nw=0,Nh=0](node14)(83,100){}
\node[Nframe=n,Nw=0,Nh=0](node15)(83,75){}
\node[Nframe=n,Nw=0,Nh=0](node16)(83,50){}
\node[Nframe=n,Nw=0,Nh=0](node17)(83,25){}
\node[Nframe=n,Nw=0,Nh=0](node18)(83,0){}

\drawedge[AHnb=0](node1,node13){}
\drawedge[AHnb=0](node2,node14){}
\drawedge[AHnb=0](node3,node15){}
\drawedge[AHnb=0](node4,node16){}
\drawedge[AHnb=0](node5,node17){}
\drawedge[AHnb=0](node6,node18){}

\drawedge[AHnb=0](node12,node11){}
\drawedge[AHnb=0](node11,node10){}
\drawedge[AHnb=0](node10,node9){}
\drawedge[AHnb=0](node9,node8){}
\drawedge(node8,node7){}
\end{picture}
\end{center}
\caption{Complexity of nonemptiness for classes of register automata}
\label{figure-ra}
\end{figure}

The results on complexity of nonemptiness for register automata
in Sections \ref{section-upper-bounds} and \ref{section-lower-bounds},
except $\Sigma^0_2$-membership of infinitary nonemptiness for 2NRA,
are depicted in Figure~\ref{figure-ra}.
The edges indicate the syntactic inclusions between automata classes.

\begin{acks}
We are grateful to
Claire David,
Massimo Franceschet,
Marcin Jurdzi\'nski,
Anca Muscholl,
David Nowak,
Jo\"el Ouaknine,
Philippe Schnoebelen and
Luc Segoufin
for helpful discussions.
\end{acks}

\bibliographystyle{acmtrans}
\bibliography{freeze}

\begin{received}
Received October 2006;
revised February 2008;
accepted April 2008
\end{received}

\end{document}